\documentclass[preprint,10pt]{elsarticle}

\usepackage{amssymb}
\usepackage{amsmath}

\newtheorem{as}{Assumption}[section]
\newtheorem{thm}{Theorem}[section]
\newtheorem{lem}{Lemma}[section]
\newtheorem{cor}{Corollary}[section]

\newtheorem{exm}{Example}[section]
\makeatletter

\@addtoreset{equation}{section}

\def\hbbe{{\widehat{\bbe}}}

\def\btheta{{\bth}}

\def\pd{\partial}
\def\hbta{\widehat{\bta}}

\def\hxi{\widehat{\xi}}
\def\MSE{{\rm MSE}}
\def\CMSE{{\rm CMSE}}
\def\Var{{\rm Var}}
\def\Cov{{\rm Cov}}

\def\hnu{\widehat{\nu}}
\def\mh{{\widehat m}}
\def\col{{\bf col}}
\def\al{{\alpha}}

\def\ep{{\varepsilon}}
\def\la{{\lambda}}

\def\th{{\theta}}

\def\bbe{{\text{\boldmath $\beta$}}}

\def\bta{{\text{\boldmath $\eta$}}}

\def\bth{{\text{\boldmath $\theta$}}}

\def\nuh{{\hat \nu}}

\def\bbeh{{\widehat \bbe}}

\def\btah{{\widehat \bta}}

\def\Si{{\Sigma}}
\def\bSi{{\text{\boldmath $\Si$}}}
\def\0{{\text{\boldmath $0$}}}
\def\a{{\text{\boldmath $a$}}}
\def\b{{\text{\boldmath $b$}}}

\def\e{{\text{\boldmath $e$}}}

\def\g{{\text{\boldmath $g$}}}

\def\r{{\text{\boldmath $r$}}}
\def\s{{\text{\boldmath $s$}}}
\def\t{{\text{\boldmath $t$}}}

\def\x{{\text{\boldmath $x$}}}
\def\y{{\text{\boldmath $y$}}}

\def\D{{\text{\boldmath $D$}}}

\def\I{{\text{\boldmath $I$}}}

\def\P{{\text{\boldmath $P$}}}

\def\R{{\text{\boldmath $R$}}}

\def\T{{\text{\boldmath $T$}}}
\def\U{{\text{\boldmath $U$}}}

\def\W{{\text{\boldmath $W$}}}

\def\Y{{\text{\boldmath $Y$}}}
\def\Z{{\text{\boldmath $Z$}}}
\def\Nc{{\cal N}}
\def\tr{{\rm tr}}

\def\vec{{\bf vec\,}}

\def\non{{\nonumber}}
\def\et{{\rm et al.}}
\def\dd{{\rm d}}
\def\zero{{\bf\text{\boldmath $0$}}}
\begin{document}

\begin{frontmatter}

\title{On Conditional Prediction Errors in Mixed Models with Application to Small Area Estimation}

\author{Shonosuke Sugasawa} 
\ead{shonosuke622@gmail.com}
\address{Graduate School of Economics, University of Tokyo, 7-3-1 Hongo, Bunkyo-ku, Tokyo 113-0033, Japan.}

\author{Tatsuya Kubokawa}
\ead{tatsuya@e.u-tokyo.ac.jp}
\address{Faculty of Economics, University of Tokyo, 7-3-1 Hongo, Bunkyo-ku, Tokyo 113-0033, Japan.}

\begin{abstract}
The empirical Bayes estimators in mixed models are useful for small area estimation in the sense of increasing precision of prediction for small area means, and one wants to know the prediction errors of the empirical Bayes estimators based on the data.
This paper is concerned with conditional prediction errors in the mixed models instead of conventional unconditional prediction errors. 
In the mixed models based on natural exponential families with quadratic variance functions, it is shown that the difference between the conditional and unconditional prediction errors is significant under distributions far from normality.
Especially for the binomial-beta mixed and the Poisson-gamma mixed models, the leading terms in the conditional prediction errors are, respectively, a quadratic concave function and an increasing function of the direct estimate in the small area, while the corresponding leading terms in the unconditional prediction errors are constants.
Second-order unbiased estimators of the conditional prediction errors are also derived and their performances are examined through simulation and empirical studies.
\end{abstract}


\end{frontmatter}

\section{Introduction}\label{sec:int}

The empirical best linear unbiased predictors (EBLUP) or empirical Bayes estimators (EB) in the Bayesian context have been used for providing reliable small-area estimates in the normal linear mixed models.
The unconditional mean squared errors (MSE) have been widely used as a measure for prediction error of EBLUP, and the asymptotic approximations of the MSEs and their approximated unbiased estimators have been studied in a lot of papers under the assumption that the number of small areas is large. 
For example, see Prasad and Rao (1990), Ghosh and Rao (1994), Rao (2003), Datta, Rao and Smith (2005) and Hall and Maiti (2006).

When data from the small area of interest are observed,  the practitioners want to know how large prediction errors the EBLUP based on the observed data have.
Concerning this issue, the conventional unconditional MSEs do not give us appropriate estimation errors, since it is an integrated measure.
Booth and Hobert (1998) suggested the conditional MSE given the data of the small area of interest, and Datta, Kubokawa, Molina and Rao (2011) and Torabi and Rao (2013) derived second-order unbiased estimators of the conditional MSE in the Fay-Herriot and nested error regression models which are well-known normal linear mixed models.
As pointed out in both papers, the difference between the conditional and unconditional MSEs is small  in the normal linear mixed models, since it appears in the second-order terms.
In the generalized linear mixed models (GLMM), however, Booth and Hobert (1998) showed that the difference is significant for distributions far from normality, since it appears in the first-order or leading terms.

Although the GLMMs are useful for analyzing count data in small area estimation, it is computationally hard to derive the EBLUP and to evaluate their conditional MSEs, because the marginal likelihood and EBLUP in the GLMM cannot be expressed in closed forms.
In fact, we need relatively high dimensional numerical integration to evaluate the conditional MSEs.
Another point is the assumption that sample sizes of small areas are large, under which the Laplace approximation can be used to get asymptotically unbiased estimators of the conditional MSEs.
However, this assumption is against the situation in small area estimation with small samples sizes.

An alternative model is the mixed model based on the natural exponential families with quadratic variance functions (NEF-QVF) suggested in Ghosh and Maiti (2004, 2008).
In the NEF-QVF mixed models, the BLUP or the Bayes estimator can be expressed explicitly as the weighed average of a sample mean and a prior mean. 
Moreover, the MSE of the empirical Bayes estimator can be approximated analytically, and their asymptotically unbiased estimator can be obtained without assuming that samples of small areas are large.
The NEF-QVF mixed models include the binomial-beta mixed and the Poisson-gamma mixed models, which are practically useful for analyzing mortality data in small areas.

Thus, in this paper, we treat the NEF-QVF mixed models instead of the GLMM and focus on the conditional prediction errors or the conditional MSEs (CMSE) of the empirical Bayes estimators (EB).
Assuming that the number of small areas is large, but sample sizes in small areas are bounded,  we not only derive second-order approximations of the conditional MSEs and their second-order unbiased estimators in closed forms, but also show that the difference between the conditional and unconditional MSEs is significant and appears in the first-order terms under distributions far from normality.

The paper is organized as follows: 
In Section \ref{sec:GMM}, the CMSE of EB is addressed in the general mixed models, and the second-order approximation of the CMSE is derived under suitable conditions on estimators of model parameters and predictors.
Second-order unbiased estimators of the CMSE are obtained in two ways of the analytical and parametric bootstrap methods.

In Section \ref{sec:NEF}, the NEF-QVF mixed models are investigated as an application of the general results in Section \ref{sec:GMM}.
The second-order approximations of the CMSEs and their second-order unbiased estimators are obtained in analytical and closed forms without assuming that sample sizes of small areas tend to infinity.
Ghosh and Maiti (2004) derived the unconditional MSE of EB, and their estimation method and techniques for analysis are heavily used in Section \ref{sec:NEF}.
It is interesting to point out that the first-order term in the CMSE is an increasing function of the direct estimate in the small area for the Poisson-gamma mixed model, and it is a quadratic concave function for the binomial-beta mixed model, while the corresponding first-order terms in the unconditional MSEs are constants for both mixed models.

Simulation and empirical studies of the suggested procedures are given in Section \ref{sec:NS}.
Two data sets are used for the empirical studies.
One is the Stomach Cancer Mortality Data in Saitama Prefecture in Japan, and the Poisson-gamma mixed model is applied.
The other is the Infant Mortality Data Before World War II in Ishikawa Prefecture in Japan, and we use the binomial-beta mixed model.
Through these analysis, it is observed that the estimates of the conditional MSEs are more variable than those of the unconditional MSEs, since conditional MSE depends on the data of the area of interest. 
For some areas, the conditional MSE gives much higher risks than the unconditional MSE, namely, the conventional MSE seems to under-estimate the conditional MSE.
Thus, we suggest providing estimates of the conditional MSE.

Finally,  the concluding remarks are given in Section \ref{sec:CR}, and the technical proofs are given in the Appendix.

\section{Conditional MSE of Empirical Bayes Estimator in General Mixed Models}
\label{sec:GMM}

Let $y=(y_1,\ldots,y_m)^t$ be a vector of observable random variables, and let $\btheta=(\theta_1,\ldots,\theta_m)^t$ be a vector of unobservable random variables. Let $\bta$ be a $q$-dimensional vector of unknown parameters.
In this paper, we treat continuous or discrete cases for $y_i$ and $\btheta$.
The conditional probability density (or mass) function of $y_i$ given $(\theta_i, \bta)$ is denoted by $f(y_i|\theta_i,\bta)$, and the conditional probability density (or mass) function of $\theta_i$ given $\bta$ is denoted by $\pi(\theta_i|\bta)$, namely,
\begin{equation}
\label{general.model}
\begin{split}
y_i | (\theta_i, \bta) \sim& f(y_i | \theta_i,\bta)
\\
\theta_i | \bta \sim& \pi(\theta_i | \bta)
\end{split}
\ \ \ \ \ \ \  i=1,\ldots,m.
\end{equation}
This expresses the general parametric mixed models. Since it can be interpreted as a Bayesian model, we here use the terminology used in Bayes statistics.
In the continuous case, the marginal density function of $y_i$ for given $\bta$ and the conditional (or posterior) density function of $\theta_i$ given $(y_i,\bta)$ are given by
\begin{equation}
\label{post}
\begin{split}
m_\pi(y_i | \bta) =& \int f(y_i | \theta_i,\bta) \pi(\theta_i|\bta) \dd \theta_i
\\
\pi(\theta_i |y_i, \bta) =& f(y_i | \theta_i,\bta) \pi(\theta_i|\bta)/m_\pi(y_i | \bta)
\end{split}
\ \ \ \ \ \ \  i=1,\ldots,m,
\end{equation}
and we use the same notations in the discrete case.
Then, for $i=1,\ldots,m$, we consider the problem of predicting a scalar quantity $\xi_i(\theta_i,\bta)$ of each small area.

When $\xi_i(\th_i,\bta)$ is predicted with $\hxi_i=\hxi_i(\y)$, the predictor $\hxi_i$ can be evaluated with the unconditional and conditional MSEs, described as
\begin{align*}
\MSE(\bta, \hxi_i)=& E\Bigl[ \bigl\{ \hxi_i - \xi_i(\theta_i,\bta) \bigr\}^2\Bigr],
\\
\CMSE(\bta, \hxi_i |y_i)=& E\Bigl[ \bigl\{ \hxi_i - \xi_i(\theta_i,\bta) \bigr\}^2|y_i\Bigr],
\end{align*}
which are denoted by MSE and CMSE, respectively.
The best predictors of $\xi_i(\th_i,\bta)$ in terms of the two kinds of MSEs are the conditional mean given by
$$
\hxi_i(y_i,\bta)= E\left[\xi_i(\theta_i,\bta)| y_i\right],
$$
which is the Bayes estimator in the Bayesian context.
Since $\bta$ is unknown, we need to estimate $\bta$ from observations $y_1,\ldots,y_m$.
Substituting an estimator $\btah$ into $\hxi_i(y_i,\bta)$ results in the empirical Bayes (EB) estimator $\hxi_i(y_i,\hbta)$.

\medskip
In this paper, we focus on asymptotic evaluations of the CMSE.
To this end, we assume the following conditions on the estimator $\btah$ and the predictor $\hxi_i(y_i,\bta)$ for large $m$:

\begin{as}\label{as:est}
\item[(i)]
 The dimension $q$ of $\bta$ is bounded and the estimator $\btah$ satisfies that $(\btah-\bta)|y_i=O_p(m^{-1/2})$, $E[\btah-\bta|y_i]=O_p(m^{-1})$ and ${\rm Var}(\hbta|y_i)=O_p(m^{-1})$ for $i=1,\ldots,m$.

\item[(ii)]
For $i=1, \ldots, m$, $\xi_i(\th_i,\bta)=O_p(1)$, $\hxi_i(y_i,\bta)=O_p(1)$, and the conditional variances of $\xi_i(\th_i,\bta)$ and $\hxi_i(y_i, \bta)$ exist.
For $j=1, \ldots, q$, the estimator $\hxi_i(y_i,\bta)$ is continuously differentiable with respect to $\eta_j$, and
$$
\partial\hxi_i(y_i,\bta)/\partial \eta_j = O_p(1),\ E[|\partial\hxi_i(y_i,\bta)/\partial \eta_j|\mid y_i]<\infty.
$$
\end{as}

Under Assumption \ref{as:est}, we get a second-order approximation of CMSE of $\hxi_i(y_i,\hbta)$.
Let
\begin{align}
T_{1i}(y_i,\bta) =& \Var(\xi_i(\theta_i,\bta) | y_i),
\label{eqn:g1}
\\
T_{2i}(y_i,\bta)=& E\Bigl[ \Bigl\{ (\btah - \bta)^t {\partial \hxi_i(y_i,\bta)\over \partial \bta}\Bigr\}^2\Bigr|y_i\Bigr],
\label{eqn:g2}
\end{align}
where $T_{1i}(y_i,\bta)$ is the conditional or posterior variance of $\xi_i(\theta_i,\bta)$.
It is noted that $T_{1i}(y_i,\bta)=O_p(1)$ and $T_{2i}(y_i,\bta)=O_p(m^{-1})$ under Assumption \ref{as:est}.

\begin{thm}
\label{thm:CMSE}
Under Assumption \ref{as:est}, the conditional MSE of $\hxi_i(y_i,\hbta)$ is approximated as
\begin{equation}
\CMSE(\bta, \hxi_i(y_i,\hbta) |y_i) = T_{1i}(y_i,\bta) + T_{2i}(y_i,\bta) + o_p(m^{-1}).
\label{eqn:gmse}
\end{equation}
\end{thm}

\noindent
{\it Proof.}\ \ 
Since $E[\xi_i - \hxi_i(y_i,\bta) | y_i] =0$, it is observed that
\begin{align}
\CMSE&(\bta, \hxi_i(y_i,\hbta) |y_i)\non\\
=& E[ \{ \xi_i(\theta_i,\bta)- \hxi_i(y_i,\bta) + \hxi_i(y_i,\bta)-\hxi_i(y_i,\hbta) \}^2|y_i]
\non\\
=&E[ \{ \xi_i(\theta_i,\bta)- \hxi_i(y_i,\bta) \}^2|y_i] + E[\{ \hxi_i(y_i,\bta)-\hxi_i(y_i,\hbta) \}^2|y_i],
\label{decom}
\end{align}
and that $E[ \{ \xi_i(\theta_i,\bta)- \hxi_i(y_i,\bta) \}^2|y_i] =Var(\xi(\theta_i,\bta)|y_i)=T_{1i}(y_i,\bta)$.
It is noted that
\begin{equation*}
\hxi_i(y_i,\hbta) = \hxi_i(y_i,\bta) + \Bigl({\pd \hxi_i(y_i,\bta^*) \over \pd\bta}\Bigr)^t (\btah-\bta),
\end{equation*}
where $\bta^*$ is between $\bta$ and $\btah$.
Since $(\btah-\bta)\mid y_i =O_p(m^{-1/2})$, we obtain
$$
E[\{ \hxi_i(y_i,\bta)-\hxi_i(y_i,\hbta) \}^2|y_i]=E\Bigl[ \Bigl\{ (\btah-\bta)^t {\partial \hxi_i(y_i,\bta)\over \partial \bta}\Bigr\}^2\Bigr|y_i\Bigr] + o_p(m^{-1}),
$$
which shows Theorem \ref{thm:CMSE}.
\hfill$\Box$

\bigskip
We next derive second-order unbiased estimators of $T_1$ and $T_2$, which result in a second-order unbiased estimator of CMSE.
As seen from Theorem \ref{thm:CMSE}, the order of $T_{2i}(y_i,\bta)$ is $O_p(m^{-1})$, so that we can estimate $T_{2i}(y_i,\bta)$ by $T_{2i}(y_i,\hbta)$ unbiasedly up to second-order. 
For estimation of $T_{1i}(y_i,\bta)$, the naive estimator $T_{1i}(y_i,\hbta)$ has a second-order bias because $T_{1i}(y_i,\bta)=O_p(1)$. 
It is observed that
\begin{equation}
E[T_{1i}(y_i,\hbta)|y_i] = T_{1i}(y_i,\bta) + T_{11i}(y_i,\bta) +T_{12i}(y_i,\bta) + o_p(m^{-1}),
\label{T1}
\end{equation}
where
\begin{equation}
\label{T11}
T_{11i}(y_i,\bta)=\Bigl(\frac{\pd T_{1i}(y_i,\bta)}{\pd \bta}\Bigr)^tE[(\hbta-\bta)|y_i]
\end{equation}
and
\begin{equation}
\label{T12}
T_{12i}(y_i,\bta)=\frac12 \tr\Bigl[ \Bigl(\frac{\pd^2 T_{1i}(y_i,\bta)}{\pd \bta\pd\bta^t}\Bigr)E\left[(\hbta-\bta)(\hbta-\bta)^t|y_i\right]\Bigr].
\end{equation}
It is noted that $T_{11i}(y_i,\bta)=O_p(m^{-1})$ and $T_{12i}(y_i,\bta)=O_p(m^{-1})$ under Assumption \ref{as:est}.

\ \\
{\bf [Analytical method]}\ \ It follows from (\ref{T1}) that a second-order unbiased estimator of CMSE is given by
\begin{equation}
\label{ana}
\widehat{\CMSE}_i(\hxi_i(y_i,\hbta))=T_{1i}(y_i,\hbta)-T_{11i}(y_i,\hbta)-T_{12i}(y_i,\hbta)+T_{2i}(y_i,\hbta).
\end{equation}

\begin{thm}
\label{thm:cMSEu1}
Under Assumption \ref{as:est}, the estimator $(\ref{ana})$ is a second-order unbiased estimator of CMSE, namely 
$$
E\bigl[\widehat{\CMSE}_i(\hxi_i(y_i,\hbta)) | y_i\bigr]=\CMSE(\bta,\hxi_i(y_i,\hbta)|y_i)+o_p(m^{-1}).
$$
\end{thm}

As explained in Section \ref{sec:NEF}, in the mixed model based on NEF-QVF, we can provide analytical expressions for $T_{11i}$ and $T_{12i}$, whereby we obtain a second-order unbiased estimator in a closed form. 
In general, however, it is hard to get analytical expressions for $T_{11i}$ and $T_{12i}$.
In this case, as given below, the parametric bootstrap method helps us provide a feasible second-order unbiased estimator of CMSE.

\ \\
{\bf [Parametric bootstrap method]}\ \ 
Since $y_i$ is fixed, a bootstrap sample is generated from
\begin{equation*}
y_j^{\ast}|(\theta_j^{\ast},\hbta)\sim f(y_j^{\ast}|\theta_j^{\ast},\hbta)\ \ \ \ \ \ \  j\neq i, \ j=1,\ldots,m,
\end{equation*}
where $\theta_j^{\ast}$'s are mutually independently distributed as $\theta_j^{\ast}|\hbta\sim \pi(\theta_j^{\ast}|\hbta)$.
Noting that $y_i$ is fixed, we construct the estimator $\hbta^{\ast}_{(i)}$ from the bootstrap sample
\begin{equation}
\label{boots}
y_1^{\ast},\ldots,y_{i-1}^{\ast},y_i,y_{i+1}^{\ast},\ldots,y_m^{\ast}
\end{equation}
with the same technique as used to obtain the estimator $\hbta$. 
Let $E_{\ast}\left[\cdot|y_i\right]$ be the expectation with regard to the bootstrap sample (\ref{boots}).
A second-order unbiased estimator of $T_{1i}(y_i,\bta)$ is given by
$$
\overline{T}_{1i}(y_i,\hbta)=2T_{1i}(y_i,\hbta)-E_{\ast}\left[T_{1i}(y_i,\hbta^{\ast}_{(i)})|y_i\right].
$$
Then, it can be verified that $E[\overline{T}_{1i}(y_i,\hbta)|y_i]=T_{1i}(y_i,\bta)+o_p(m^{-1})$.
In fact, from (\ref{T1}), it is noted that
$$
E[T_{1i}(y_i,\hbta)|y_i]=T_{1i}(y_i,\bta)+d_i(y_i,\bta)+o_p(m^{-1}),
$$
where $d_i(y_i,\bta) = T_{11i}(y_i,\bta) +T_{12i}(y_i,\bta)$.
This implies that $E_{\ast}\left[T_{1i}(y_i,\hbta^{\ast}_{(i)})|y_i\right]=T_{1i}(y_i,\hbta)+d_i(y_i,\hbta)+o_p(m^{-1})$.
Since $d_i(y_i,\bta)$ is continuous in $\bta$ and $d_i(y_i,\bta)=O_p(m^{-1})$, one gets $E[\overline{T}_{1i}(y_i,\hbta)|y_i]=T_{1i}(y_i,\bta)+o_p(m^{-1})$.

For $T_{2i}(y_i,\bta)$, from (\ref{decom}), it is estimated via parametric bootstrap method as
$$
T_{2i}^\ast(y_i,\hbta)=
E^{\ast}\bigl[ \{ \hxi_i^{\ast}(y_i,\hbta)-\hxi_i^{\ast}(y_i,\hbta_{(i)}^{\ast}) \}^2 \bigr|y_i \bigr].
$$
It is noted that the estimator $T_{2i}^\ast(y_i,\hbta)$ is always available although an analytical expression of $T_{2i}(y_i,\bta)$ is not necessarily available. 
Combining the above results yields the estimator
\begin{equation}
\label{boot}
\widehat{\CMSE}_i^{\ast}(\hxi_i(y_i,\hbta))=\overline{T}_{1i}(y_i,\hbta)+T_{2i}^\ast(y_i,\hbta).\end{equation}

\begin{thm}
\label{thm:cMSEu2}
Under Assumption \ref{as:est}, the estimator $(\ref{boot})$ is a second-order unbiased estimator of CMSE, namely
$$
E[\widehat{\CMSE}_i^{\ast}|y_i]=\CMSE(\bta, \hxi_i(y_i,\hbta)|y_i)+o_p(m^{-1}).
$$
\end{thm}

\section{Applications to NEF-QVF}
\label{sec:NEF}

We now consider the mixed models based on natural exponential families with quadratic variance functions (NEF-QVF). 
The NEF-QVF mixed models were used in context of small area estimation by Ghosh and Maiti (2004), who evaluated asymptotically the unconditional MSE for calibrating uncertainty of the empirical Bayes estimator when $m$ is large. 
In this section, we handle an area level model with a survey estimate from each area where the survey estimate has a distribution based on NEF-QVF, and apply the results in the previous section to provide a second-order approximation and its unbiased estimator for the conditional MSE of the EB. 

In our settings, it is assumed that known parameters $n_i$'s, which correspond to sample sizes in small-areas in normal cases, are bounded and the number of areas $m$ is large.

\subsection{Empirical Bayes estimator in NEF-QVF}
Let $y_1, \ldots, y_m$ be mutually independent random variables where the conditional distribution of $y_i$ given $\th_i$ and the marginal distribution of $\th_i$ belong to the the following natural exponential families:
\begin{equation}
\begin{split}
y_i | \th_i \sim& f(y_i|\th_i)=\exp[ n_i(\th_i y_i -\psi(\th_i)) + c(y_i,n_i)],
\\
\th_i | \nu, m_i\sim& \pi(\th_i|\nu, m_i) = \exp[\nu(m_i\th_i-\psi(\th_i))]C(\nu,m_i),
\end{split}
\label{model}
\end{equation}
where $n_i$ is a known scalar parameter and $\nu$ is an unknown scalar hyperparameter.
Let $\y=(y_1, \ldots, y_m)^t$ and $\bth=(\th_1, \ldots, \th_m)^t$.
The function $f(y_i|\th_i)$ is the regular one-parameter exponential family and the function $\pi(\th_i|\nu, m_i)$ is the conjugate prior distribution.
Define $\xi_i$ by
$$
\xi_i = E[y_i|\th_i] = \psi'(\th_i),
$$
which is the conditional expectation of $y_i$ given $\th_i$, where $\psi'(x)=d \psi(x)/d x$.
Assume that $\psi''(\th_i)=Q(\xi_i)$ for $\psi''(x)=d^2 \psi(x)/d x^2$, namely, 
$$
\Var(y_i|\th_i)={\psi''(\th_i)\over n_i}={Q(\xi_i)\over n_i},
$$
where $Q(x) = v_0 + v_1 x +v_2 x^2$ for known constants $v_0$, $v_1$ and $v_2$ which are not simultaneously zero.
This means that given $\th_i$, the conditional variance $Var(y_i|\theta_i)$ is a quadratic function of the conditional expectation $E[y_i|\th_i]$.
This is the natural exponential family with the quadratic variance function (NEF-QVF) studied by Morris (1982, 1983).
Similarly, the mean and variance of the prior distribution are given by
\begin{equation}
E[\xi_i | m_i, \nu] =m_i, \quad \Var(\xi_i|m_i,\nu)={Q_i(m_i)\over \nu - v_2}.
\label{eqn:emv}
\end{equation}
In our settings, we consider the link given by 
$$
m_i=\psi'(\x_i^t\bbe), \quad i=1, \ldots, m,
$$
where $\x_i$ is a $p\times 1$ vector of explanatory variables and $\bbe$ is a $p\times 1$ unknown common vector of regression coefficients.
Then, the unknown parameters $\bta$ in the previous section correspond to $\bta^t=(\bbe^t, \nu)$.
The joint probability density (or mass) function of $(y_i, \th_i)$ can be expressed as
$$
f(y_i|\th_i)\pi(\th_i|\nu, m_i) = \pi(\th_i|y_i, \nu) f_\pi(y_i|\nu, m_i),
$$
where $\pi(\th_i|y_i, \nu)$ is the conditional (or posterior) density function of $\th_i$ given $y_i$, and $f_\pi(y_i|\nu, m_i)$ is the marginal density function of $y_i$.
These density (or mass) functions are written as
\begin{equation}
\begin{split}
\pi(\th_i|y_i, \nu, m_i) =& \exp[ (n_i+\nu)(\hxi_i \th_i - \psi(\th_i))] C(n_i+\nu, \hxi_i),
\\
f_\pi(y_i|\nu, m_i) =& {C(\nu,m_i)\over C(n_i+\nu,\hxi_i)} \exp[ c(y_i, n_i)],
\end{split}
\label{eqn:ecmodel}
\end{equation}
where $\hxi_i$ is the posterior expectation of $\xi_i$, namely, $\hxi_i=E[\xi_i|y_i, \bta]$, given by
\begin{equation}
\hxi_i=\hxi_i(y_i,\bta) ={n_i y_i + \nu m_i \over n_i + \nu},
\label{B}
\end{equation}
which corresponds to the Bayes estimator of $\xi_i$ in the Bayesian context when $\nu$ and $m_i$ are known.
As shown in Ghosh and Maiti (2004), 
\begin{align*}
E[y_i]=&E[\psi'(\th_i)]=m_i,
\\
\Var(y_i)=&\Var(E[y_i|\th_i])+E[\Var(y_i|\th_i)]=\Var(\xi_i)+E[Q_i(\xi_i)/n_i]=Q_i(m_i)\phi_i,
\\
\Cov(y_i, \xi_i)=&E[\Cov(y_i,\xi_i)|\th_i]+ \Cov(E[y_i|\th_i], \xi_i)=Q_i(m_i)/(\nu-v_2),
\end{align*}
for $\phi_i=(1+\nu/n_i)/(\nu-v_2)$.
Using these observations, Ghosh and Maiti (2004) showed that the Bayes estimator $\hxi_i$ given in (\ref{B}) is the best linear unbiased predictor (BLUP) of $\xi_i$ in terms of MSE.

Since the hyperparameters $\bta$ are unknown, we need to estimate them from the joint marginal distribution of $\y$. 
For the purpose, Ghosh and Maiti (2004) suggested the estimating equations given in Godambe and Thompson (1989).
Let $\g_i=(g_{1i}, g_{2i})^t$ for $g_{1i}=y_i-m_i$ and $g_{2i}=(y_i-m_i)^2-\phi_iQ_i(m_i)$.
Let
\begin{align*}
\D_i^t=& Q_i(m_i) \left(\begin{array}{cc} \x_i& Q'_i(m_i)\phi_i\x_i\\ 0 & -(1+v_2/n_i)(\nu-v_2)^{-2}\end{array} \right),
\\
\bSi_i =& \Cov(\g_i) = \left(\begin{array}{cc} \mu_{2i} & \mu_{3i}\\ \mu_{3i}& \mu_{4i}-\mu_{2i}^2 \end{array} \right),
\end{align*}
and $|\bSi_i|=\mu_{4i}\mu_{2i}-\mu_{2i}^3-\mu_{3i}^2$, where $\mu_{ri}=E[(y_i-m_i)^r]$, $r=1, 2, \ldots$, and exact expressions of $\mu_{2i}$, $\mu_{3i}$ and $\mu_{4i}$ are given below.
Then, Ghosh and Maiti (2004) derived the estimating equations given by $\sum_{i=1}^m \D_i^t\bSi_i^{-1}\g_i=\zero$, which are written as
\begin{equation}
\begin{split}
\sum_{i=1}^m{1\over |\bSi_i|} \Bigl[ &\{\mu_{4i}-\mu_{2i}^2-\mu_{3i}\phi_i Q'_i(m_i)\}g_{1i}
 + \{\mu_{2i}\phi_iQ'_i(m_i) - \mu_{3i}\}g_{2i}\Bigr] Q_i(m_i) \x_i =\zero,
\\
\sum_{i=1}^m {1\over |\bSi_i|} &\{\mu_{2i}g_{2i}-\mu_{3i}g_{1i}\}Q_i(m_i)(1+v_2/n_i)(\nu-v_2)^{-2}=0.
\end{split}
\label{ee}
\end{equation}
The resulting estimator of $\bta$ is here called thel GT-estimator and denoted by $\btah_{\rm GT}$.
The equations can be solved numerically. 
In our numerical investigation, we used the {\tt optim} function in ^^ R' to solve the estimating equations by minimizing the sums of squares of the estimating functions.
This approach may cause the problem in the presence of multiple roots, but fortunately we did not encounter this situation in our examples given in Section \ref{sec:EE}. 

The exact moments $\mu_{ri}=E[(y_i-m_i)^r]$, $r=1, 2, 3, 4$, are obtain from Theorem 1 of Ghosh and Maiti (2004) as
\begin{align*}
\mu_{2i}=&{Q(m_i)( \nu/n_i +1) \over \nu-v_2}, \quad 
\mu_{3i}={Q(m_i)Q'(m_i)( \nu/n_i+1)( \nu/n_i+2) \over (\nu-v_2)(\nu-2v_2)},
\end{align*}
and
\begin{align*}
\mu_{4i}=& ( d_i+1)(2 d_i+1)(3 d_i+1)E[(\xi_i-m_i)^4]
+ {6\over n_i}Q_i'(m_i)( d_i+1)(2 d_i+1)E[(\xi_i-m_i)^3]
\\
&+{ d_i + 1\over n_i^2}\bigl[ 7 \{Q'(m_i)\}^2+2n_i(4 d_i+3)Q(m_i)\bigr] E[(\xi_i-m_i)^2] 
\\
&+ {1\over n_i^3}Q(m_i)\bigl[ n_i(2 d_i+3)Q(m_i)+\{Q'(m_i)\}^2\bigr],
\end{align*}
for $ d_i=v_{2}/n_i$. 
The expressions of the moments of $\xi_i$ are obtained given in Kubokawa, $\et$ (2014) as $E[(\xi_i-m_i)^2]=Q(m_i)/(\nu-v_2)$, $E[(\xi_i-m_i)^3]=2Q(m_i)Q'(m_i)/(\nu-v_2)(\nu-2v_2)$ and
$$
E\left[(\xi_i-m_i)^4\right]=\frac{3Q(m_i)\left[(\nu-v_2)Q(m_i)+2\left\{Q'(m_i)\right\}^2\right]}{(\nu-v_2)(\nu-2v_2)(\nu-3v_2)}.
$$
Using these expressions, we obtain the GT-estimator $\hbta=(\hbbe^t,\hnu)^t$.

An alternative method for estimating $\bta$ is the maximum likelihood estimator (ML).
Since a closed expression of the marginal distribution of $\y$ is given in (\ref{eqn:ecmodel}) in the NEF-QVF mixed model, the ML-estimator of $\bta$ is provided by
\begin{equation}\label{estML}
\hbta_{\rm ML}={\rm argmax}_{\bta}\left\{\sum_{i=1}^m\log \frac{C(\nu,m_i)}{C(n_i+\nu,\hxi_i(y_i,\bta))}\right\}.
\end{equation}
Since we do not have a closed expression of the maximizer, we resort to a numerical optimization. 

When the parameter $\bta$ is estimated by the GT-estimator $\hbta=\hbta_{\rm GT}$ or the ML-estimator $\hbta=\hbta_{\rm ML}$, we can construct the estimator $\mh_i=\psi'(\x_i^t\bbeh)$ for $m_i$.
Substituting $\mh_i$ and $\nuh$ into (\ref{B}), we finally get the empirical Bayes estimator of $\xi_i$, given by
\begin{equation}
\hxi_i(y_i, \btah) = {n_i y_i + \hnu \mh_i \over n_i + \hnu}.
\label{EB}
\end{equation}
The EB estimator is often used as a predictor in small area estimation and its uncertainty is of great importance. 
Our interest is in evaluation of the conditional MSE of $\hxi_i(y_i, \btah)$, which is investigated in the next subsection.

\subsection{Evaluation of the conditional MSE}

Since the second-order approximation of the conditional MSE is given in Theorem \ref{thm:CMSE}, we need to evaluate the first and second order terms $T_{1i}(y_i,\bta)$ and $T_{2i}(y_i,\bta)$ in the CMSE.
For the first order term, it is easy to see that 
\begin{equation}
T_{1i}(y_i,\bta)=\Var(\xi_i(\theta_i,\bta) | y_i)=\frac{Q(\hxi_i(y_i,\bta))}{n_i+\nu-v_2}, \ \ \ \ \ i=1,\ldots,m,
\label{T1n}
\end{equation}
which is $O_p(1)$. 
For the second order term, unfortunately, we do not have an analytical expression of $T_{2i}(y_i,\bta)$ when we use the ML-estimator $\hbta_{\rm ML}$ for $\hbta$. 
But, the parametric bootstrap method given in Theorem \ref{thm:cMSEu2} enables us to construct the second-order unbiased estimator of the CMSE.
When the GT-estimator $\hbta_{\rm GT}$ is used for $\bta$, on the other hand, we can derive an analytical expression of $T_{2i}(y_i,\bta)$, which yields closed forms of the second-order approximation of the CMSE and the asymptotically unbiased estimator of the CMSE. 
Thus, in the rest of this subsection, we focus on derivation of analytical expressions for the CMSE when the GT-estimator $\hbta_{\rm GT}$ is used for $\bta$.

We begin by giving a stochastic expansion and conditional moments of $\hbta_{\rm GT}$ which is the solution of the estimating equations (\ref{ee}). 
We use the notations given by 
\begin{align*}
\s_m=\sum_{i=1}^m &\D_i^t\bSi_i^{-1}\g_i, \ \ \ \ \ \ 
\U(\bta)=\sum_{i=1}^m \D_i^t\bSi_i^{-1}\D_i \ (=\Cov(\s_m)),
\\
&\b(y_i, \bta)=\U(\bta)^{-1}\Bigl(\D_i^t\bSi_i^{-1}\g_i+\a_1(\bta)+\frac12\a_2(\bta)\Bigr),
\end{align*}
where the detailed forms of $\a_1(\bta)$ and $\a_2(\bta)$ are given in (\ref{a1}) and (\ref{a2}) in the Appendix, respectively.
It is noted that $\s_m=O_p(m^{1/2})$ and $\U(\bta)=O(m)$. 
The following lemma is useful for evaluating the conditional MSE, where the proof is given in the Appendix.

\begin{lem}
\label{lem:1}
Let $\hbta_{\rm GT}$ be the solution of estimating equations in $(\ref{ee})$.
Then for $i=1, \ldots,m$, 
\begin{equation}
\begin{split}
&(\btah_{\rm GT}-\bta)|y_i=\U(\bta)^{-1}\s_m+ o_p(m^{-1/2}), \\ 
&E[(\btah_{\rm GT}-\bta)(\btah_{\rm GT}-\bta)^t|y_i]=\U(\bta)^{-1}+o_p(m^{-1}),\\
&E[\btah_{\rm GT}-\bta|y_i]=\b (y_i,\bta)+o_p(m^{-1}).
\end{split}
\label{eqn}
\end{equation}
\end{lem}

Lemma \ref{lem:1} means that the second-order approximations of the conditional moments $E[(\btah_{\rm GT}-\bta)(\btah_{\rm GT}-\bta)^t|y_i]$ and $E[\btah_{\rm GT}-\bta|y_i]$ do not depend on $y_i$, that is, they are equal to the unconditional moments given in Ghosh and Maiti (2004).
Lemma \ref{lem:1} shows that the estimator $\hbta_{\rm GT}$ satisfies Assumption \ref{as:est}.
It is noted that the partial derivatives of $\D_i^t\bSi^{-1}$ and $\D_i$ appear in the expressions (\ref{a1}) and (\ref{a2}), and these can be numerically evaluated using the numerical derivatives, where the detailed procedure is given in the Appendix.

We now derive analytical expressions $T_{2i}(y_i,\bta)$ in Theorem \ref{thm:CMSE}. 
In the following theorem, we can evaluate $T_{2i}(y_i,\bta)$ as
\begin{equation}
T_{2i}(y_i,\bta)=\tr\left[\P_i(y_i,\bta)\U(\bta)^{-1}\right],
\label{T2n}
\end{equation}
which is $O_p(m^{-1})$, where 
$$
\P_i(y_i,\bta)=(n_i+\nu)^{-2}\left(\begin{array}{cc} \nu^2Q(m_i)^2\x_i\x_i^t& -n_i\nu(n_i+\nu)^{-1}Q(m_i)g_{1i}\x_i\\
-n_i\nu(n_i+\nu)^{-1}Q(m_i)g_{1i}\x_i^t & n_i^2(n_i+\nu)^{-2}g_{1i}^2\end{array} \right).
$$

\begin{thm}
\label{thm:cnef}
The CMSE of $\hxi(y_i,\hbta_{\rm GT})$ can be approximated up to $O_p(m^{-1})$ as
\begin{equation}
\label{CMSE}
\CMSE_i(\bta, \hxi_i(y_i,\hbta_{\rm GT})|y_i)=T_{1i}(y_i,\bta)+T_{2i}(y_i,\bta)+o_p(m^{-1}),
\end{equation}
where $T_{1i}(y_i,\bta)$ and $T_{2i}(y_i,\bta)$ are given in $(\ref{T1n})$ and $(\ref{T2n})$, respectively.
\end{thm}

\noindent
{\it Proof.}
From Theorem \ref{thm:CMSE}, it is sufficient to calculate $T_{2i}$, which is written as
\begin{align*}
E\Bigl[ \Bigl\{ (\btah_{\rm GT}-\bta)^t {\partial \hxi_i(y_i,\bta)\over \partial \bta}\Bigr\}^2\Bigr|y_i\Bigr]
&= \tr E\Bigl[\Bigl(\frac{\pd\hxi_i}{\pd\bta}\Bigr)\Bigl(\frac{\pd\hxi_i}{\pd\bta}\Bigr)^t(\hbta_{\rm GT}-\bta)(\hbta_{\rm GT}-\bta)^t\Bigr|y_i\Bigr]\\
&=\tr\Bigl[\Bigl(\frac{\pd\hxi_i}{\pd\bta}\Bigr)\Bigl(\frac{\pd\hxi_i}{\pd\bta}\Bigr)^tE\left[(\hbta_{\rm GT}-\bta)(\hbta_{\rm GT}-\bta)^t\big|y_i\right]\Bigr].
\end{align*}
It is noted from (\ref{B}) that
$$
\frac{\pd\hxi_i(y_i,\bta)}{\pd\bta}=\left(\begin{array}{c}
\nu(n_i+\nu)^{-1}Q(m_i)\x_i\\
-n_i(n_i+\nu)^{-2}g_{1i}
\end{array}\right).
$$ 
Then from Lemma \ref{lem:1}, the last formula can be approximated as
$$
\tr\left[\P_i(y_i,\bta)\U(\bta)^{-1}\right]+o_p(m^{-1}),
$$
which completes the proof.
\hfill$\Box$

\bigskip
Taking the expectation of $\CMSE_i$ with respect to $y_i$, one gets the unconditional MSE given in Theorem 1 of Ghosh and Maiti (2004) with $\delta_i=n_i^{-1}$.
In fact, 
\begin{align*}
T_{1i}(\bta)\equiv& E[T_{1i}(y_i,\bta)]=\frac{\nu}{(n_i+\nu)(\nu-v_2)}Q(m_i),
\\
T_{2i}(\bta)\equiv &E[T_{2i}(y_i,\bta)]\\
=&(n_i+\nu)^{-2}\tr\Bigl[\Bigl(\begin{array}{cc} \nu^2Q(m_i)^2\x_i\x_i^t&\0\\
\0^t & n_i(n_i+\nu)^{-1}Q(m_i)(\nu-v_2)^{-1}\end{array} \Bigr)\U(\bta)^{-1}\Bigr].
\end{align*}

\begin{cor}
The unconditional MSE of $\hxi_i(y_i,\hbta_{\rm GT})$ is approximated as
\begin{equation}
\label{uMSE}
\MSE(\bta, \hxi_i(y_i,\hbta_{\rm GT}))=T_{1i}(\bta)+T_{2i}(\bta)+o(m^{-1}).
\end{equation}
\end{cor}

It is interesting to investigate the difference between the approximations of the CMSE and the MSE.
When the underlying distribution of $y_i$ is a normal distribution, we have $Q(x)=1$, or $v_0=1$ and $v_1=v_2=0$, so that $T_{1i}(y_i,\bta)=1/(n_i+\nu)=T_{1i}(\bta)$, namely the leading term in the CMSE is identical to that in the MSE.
Thus, the difference between the CMSE and the MSE appears in the second-order term with $O_p(m^{-1})$. 
When $v_1$ or $v_2$ is not zero, however, the leading term $T_{1i}(y_i,\bta)$ in the CMSE is a function of $y_i$ and it is not equal to the leading term $T_{1i}(\bta)$ in the MSE.
Thus, for distributions far from the normality, the difference between the CMSE and the MSE is significant even when $m$ is large. 
This tells us about the remark that one cannot replace the conditional MSE given $y_i$ with the corresponding unconditional MSE except for the normal distribution.
Some examples including the Poisson and binomial distributions are given in Section \ref{sec:exm}.

We next derive an analytical form of a second-order unbiased estimator for the CMSE. 
For the purpose, we need to calculate $T_{11i}$ and $T_{12i}$ given in (\ref{T11}) and (\ref{T12}), respectively.
Note that
\begin{align*}
\r(y_i,\bta) \equiv& \frac{\pd T_{1i}}{\pd\bta}=\Bigl(\begin{array}{c}
\nu(n_i+\nu)^{-1}\la_iQ'(\hxi_i)Q(m_i)\x_i\\
-\la_i^{2}Q(\hxi_i)-\la_in_i(n_i+\nu)^{-2}Q'(\hxi_i)g_{1i}
\end{array}\Bigr),
\\
\R(y_i,\bta)\equiv& \frac{\pd^2 T_{1i}}{\pd\bta\pd\bta^t}=\Bigl(\begin{array}{cc}
\T_{1i}^{11}  & \T_{1i}^{12}\\
(\T_{1i}^{12})^t  & T_{1i}^{22}
\end{array}\Bigr),
\end{align*}
where $\la_i=(n_i+\nu-v_2)^{-1}$, and
\begin{align*}
\T_{1i}^{11}&=(n_i+\nu)^{-2}\nu\x_i\x_i^t\la_iQ(m_i)\left[2v_2\nu Q(m_i)+Q'(\hxi_i)Q'(m_i)(n_i+\nu)\right],\\
\T_{1i}^{12}&=\frac{\pd^2T_{1i}}{\pd\bbe\pd\nu}=Q(m_i)\la_i(n_i+\nu)^{-2}\left\{Q'(\hxi_i)\left(n_i-\nu(n_i+\nu)\la_i\right)-2v_2n_i\nu g_{1i}(n_i+\nu)^{-1}\right\}\x_i,\\
T_{1i}^{22}&=\frac{\pd^2T_{1i}}{\pd\nu^2}=2\la_i^3Q(\hxi_i)+2\la_i^2 n_i(n_i+\nu)^{-2}Q'(\hxi_i)g_{1i}
\\
&\hspace{3cm} 
+2\la_in_i(n_i+\nu)^{-4}g_{1i}\left[(n_i+\nu)Q'(\hxi_i)+n_i v_2g_{1i}\right].
\end{align*}
Using (\ref{eqn}) in Lemma \ref{lem:1}, we obtain the analytical expressions of $T_{11i}$ and $T_{12i}$ as
\begin{align*}
T_{11i}(y_i,\bta)=&\r(y_i,\bta)^t \b (y_i,\bta),
\\
T_{12i}(y_i,\bta)=&\frac12\tr\left[\R(y_i,\bta)\U(\bta)^{-1}\right].
\end{align*}
The estimator $\widehat{\CMSE}_i$ given in (\ref{ana}) is expressed as
\begin{align}
\label{estcmse}
\widehat{\CMSE}_i(\hxi_i(y_i,\hbta_{\rm GT}))=& T_{1i}(y_i,\hbta_{\rm GT})+T_{2i}(y_i,\hbta_{\rm GT})-\r(y_i,\hbta_{\rm GT})^t\b (y_i,\hbta_{\rm GT})
\non\\
&-\frac12\tr\left[\R(y_i,\hbta_{\rm GT})\U(\hbta_{\rm GT})^{-1}\right].
\end{align}

\begin{thm}
\label{thm:unef}
The estimator $(\ref{estcmse})$ is a second-order unbiased estimator, namely, 
$$
E[{\widehat{\CMSE}}_i(\hxi_i(y_i,\hbta_{\rm GT})) \mid y_i ]=\CMSE_i(\bta, \hxi_i(y_i,\hbta_{\rm GT}) \mid y_i)+o_p(m^{-1}).
$$
\end{thm}

It is noted that the results in Theorems \ref{thm:cnef} and \ref{thm:unef} do not require the condition that $n_i\to\infty$.
Thus, the results in Theorems \ref{thm:cnef} and \ref{thm:unef} are applicable in the context of small area estimation.

\subsection{Some useful examples}
\label{sec:exm}

We give some examples of the mixed models belonging to (\ref{model}) and investigate the conditional MSE. 

\ \\
{\bf [1] Fay-Herriot model}. The Fay-Herriot model is an area-level model often used in small area estimation, given by
$$
y_i=\x_i^t\bbe+v_i+\ep_i, \ \ \ \ i=1,\ldots,m,
$$
where $m$ is the number of small areas, and $v_i$'s and $\ep_i$'s are mutually independently distributed random errors such that $v_i\sim \Nc(0,A)$ and $\ep_i\sim\Nc(0,D_i)$. 
The notations in (\ref{model}) correspond to $n_i=D_i^{-1}, v_0=1, v_1=v_2=0, \xi_i=\theta_i, \nu=A^{-1}$ and $\psi(\theta_i)=\theta_i^2/2$. 
In this case, the estimating equations in (\ref{ee}) reduce to
\begin{equation*}
\begin{split}
&\sum_{i=1}^m(A+D_i)^{-1}\x_iy_i=\sum_{i=1}^m(A+D_i)^{-1}\x_i\x_i^t\bbe,\\
&\sum_{i=1}^m(A+D_i)^{-2}(y_i-\x_i^t\bbe)^2=\sum_{i=1}^m(A+D_i)^{-1},
\end{split}
\end{equation*}
which coincide with the likelihood equations for the maximum likelihood estimators of $\bbe$ and $A$, namely $\hbta_{\rm ML}=\hbta_{\rm GT}$ in Fay-Herriot model. 
The terms $T_{1i}(y_i,\bta)$ and $T_{2i}(y_i,\bta)$ in approximation (\ref{CMSE}) of the CMSE are written as
\begin{align*}
T_{1i}(y_i,\bta)&=\frac{AD_i}{A+D_i}\\
T_{2i}(y_i,\bta)&=\frac{D_i}{(A+D_i)^2}\x_i^t\Bigl(\sum_{j=1}^m\frac{\x_j\x_j^t}{A+D_j}\Bigr)^{-1}\x_j+\frac{D_i^2(y_i-\x_i^t\bbe)^2}{(A+D_j)^4}\Bigl(\sum_{j=1}^m\frac{1}{2(A+D_j)^2}\Bigr)^{-1},
\end{align*}
which were given in Datta \et (2011). 
In the Fay-Herriot model, $T_{1i}(y_i,\bta)={AD_i}/({A+D_i})=T_{1i}(\bta)$, namely, the leading terms in the conditional and unconditional MSEs are identical, and the difference between the CMSE and MSE is small for large $m$.

\ \\
{\bf [2] Poisson-gamma mixture model}. Let $z_1,\ldots,z_m$ be mutually independent random variables having
$$
z_i|\la_i\sim{\rm Po}(n_i\la_i) \quad {\rm and} \quad \la_i\sim{\rm Ga}(\nu m_i,1/\nu)
$$
where $\la_1,\ldots,\la_m$ are mutually independent, Po$(\la)$ denotes the Poisson distribution with mean $\la$, and Ga$(a,b)$ denotes the gamma distribution with shape parameter $a$ and scale parameter $b$. 
Let $y_i=z_i/n_i$ and $\log m_i=\x_i^t\bbe$ for $i=1,\ldots,m$. 
Then, the notations in (\ref{model}) correspond to $v_1=1, \ v_0=v_2=0, \ \xi_i=\la_i=\exp(\theta_i),$ and $\psi(\theta_i)=\exp(\theta_i)$. 
The posterior distribution of $\la_i$ is ${\rm Ga}(\nu m_i+n_iy_i,(n_i+\nu)^{-1})$ or ${\rm Ga}((n_i+\nu)\hxi_i,(n_i+\nu)^{-1})$. 
Then we have
$$
T_{1i}(y_i,\bta)=\frac{\hxi(y_i,\bta)}{n_i+\nu}=\frac{n_iy_i+\nu m_i}{(n_i+\nu)^2},
$$
which increases in $y_i$. 
Thus, the difference between the conditional and unconditional MSEs increases in $y_i$. 
When a large value of $y_i$ is observed, it should be remarked that the conditional MSE of the empirical Bayes estimator given $y_i$ is larger than the unconditional (or integrated) MSE.
Hence, it is meaningful to provide to practitioners the information on the conditional MSE as well as the unconditional MSE.

For the Poisson-gamma mixture model, the marginal distribution of $y_i$ (marginal likelihood) is the negative binomial distribution given by
$$
f(y_i|\bta)=\frac{\Gamma(n_iy_i+\nu m_i)}{\Gamma(n_iy_i+1)\Gamma(\nu m_i)}\left(\frac{n_i}{n_i+\nu}\right)^{n_iy_i}\left(\frac{\nu}{n_i+\nu}\right)^{\nu m_i},
$$ 
where $\Gamma(\cdot)$ denotes a gamma function. 
Thus it is noted that the maximum likelihood estimator can be obtained by maximizing $\sum_{i=1}^m\log f(y_i|\bta)$.

\ \\
{\bf [3] Binomial-beta mixture model.} Let $z_1,\ldots,z_m$ be mutually independent random variables having
$$
z_i|p_i\sim {\rm Bin}(n_i,p_i) \quad {\rm and}\quad p_i\sim {\rm Beta}(\nu m_i,\nu(1-m_i)),
$$
where $p_1,\ldots,p_m$ are mutually independent, Bin$(n,p)$ denotes the binomial distribution and Beta$(a,b)$ denotes the beta distribution. 
Let $y_i=z_i/n_i$ and $m_i=\exp(\x_i^t\bbe)/(1+\exp(\x_i^t\bbe))$ for $i=1,\ldots,m$. 
Then the notations in (\ref{model}) correspond to $v_0=0, \ v_1=1$ and $v_2=-1, \ \xi_i=p_i=\exp(\theta_i)/(1+\exp(\theta_i))$ and $\psi(\theta_i)=\log(1+\exp(\theta_i))$. The posterior distribution of $p_i$ is Beta$(\nu m_i+n_i y_i,n_i(1-y_i)+\nu(1-m_i))$ or Beta$((n_i+\nu)\hxi_i,(n_i+\nu)(1-\hxi_i))$, so that $T_{1i}(y_i,\bta)$ is written as
$$
T_{1i}(y_i,\bta)=\frac{\hxi_i(y_i,\bta)(1-\hxi_i(y_i,\bta))}{n_i+\nu+1},
$$
which is a quadratic and concave function of $y_i$. 
Since $0<\hxi(y_i,\bta)<1$, $T_{1i}(y_i,\bta)$ is always positive and attains the maximum when $\hxi_i=1/2$ or $y_i=(n_i+\nu)/2n_i-\nu m_i/n_i$, and $T_{1i}(y_i,\bta)=0$ when $\hxi_i=0\ {\rm or}\ 1$. 
Thus, the value of $T_{1i}(y_i,\bta)$ is relatively small when $y_i$ is close to $0$ or $1$. 
When $y_i$ is around $1/2$, the value of $T_{1i}(y_i,\bta)$ tends to be larger. 
When a value around $1/2$ is observed for $y_i$, it should be remarked that the conditional MSE of the EB given $y_i$ is larger than the unconditional (or integrated) MSE.

For the binomial-beta mixture model, the marginal likelihood is proportional to
$$
L(\bta)\propto\prod_{i=1}^m\frac{B(\nu m_i+n_i y_i, n_i(1-y_i)+\nu(1-m_i))}{B(\nu m_i,\nu(1-m_i))},
$$
where $B(\cdot)$ denotes a beta function. 
Then, the MLE of the parameters can be obtained as a maximizer of the marginal likelihood.

\section{Numerical and Empirical Studies}
\label{sec:NS}

We here give some comparisons of the conditional and unconditional MSEs and investigate finite sample performances of the second-order unbiased estimator of the CMSE.
We also apply the suggested procedures to real mortality data.

\subsection{Comparison of the conditional and unconditional MSEs}
\label{sec:com}

\begin{figure}[!thb]
\label{fig1}
\centering
\includegraphics[width=9cm,clip]{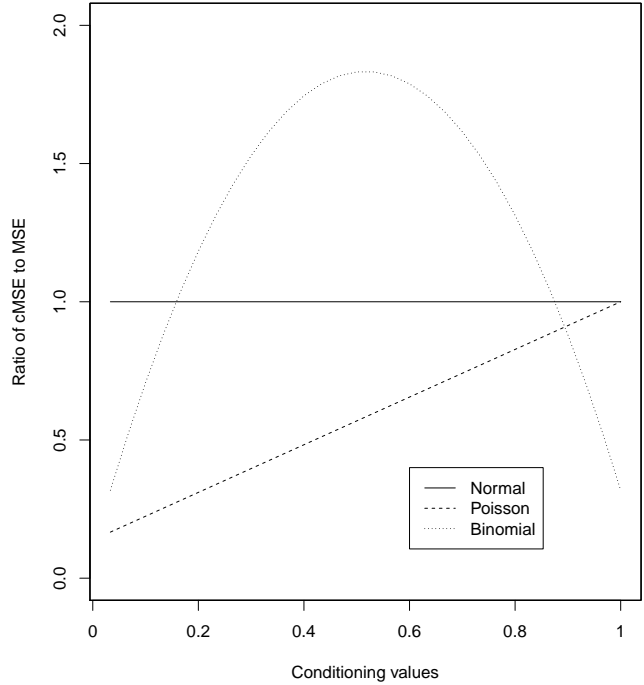}
\caption{Figures of ${\rm Ratio}_1$ for the Three Mixed Models\   (The solid, dashed and dotted lines correspond to the Fay-Herriot, Poisson-gamma mixture and binomial-beta mixture models, respectively.)
}
\label{fig:comp}
\end{figure}

It is interesting to investigate how different the conditional MSE is from the unconditional MSE.
The major difference between them appears in the leading terms, namely the terms with order $O_p(1)$ in the CMSE and MSE.
The ratio of the leading term of the CMSE to that of the MSE is defined by
$$
{\rm Ratio}_1=T_{1i}(y_i,\bta)/E[T_{1i}(y_i,\bta)],
$$
which is a function of $y_i$ and $\bta$.
We consider the case that $m=10$, $\nu=1$, $\x_i^t\bbe=\mu=0$ and $n_i=10$ for $i=1,\ldots,m$.
Then, the curves of the functions ${\rm Ratio}_1$ are illustrated Figure \ref{fig:comp} for the three mixed models: the Fay-Herriot, Poisson-gamma and binomial-beta models.
As mentioned before, in the Fay-Herriot (or normal-normal mixture) model, ${\rm Ratio}_1=1$ since $T_{1i}(y_i,\bta)=E\left[T_{1i}(y_i,\bta)\right]$.
For the Poisson-gamma and binomial-beta mixture models, Figure \ref{fig:comp} tells us about the interesting features of their leading terms in the CMSE, namely, the ratio is an increasing function of $y_i$ for the Poisson-gamma mixture model, and a concave and quadratic function of $y_i$ for the binomial-beta mixture model.

\begin{figure}[!thb]
\label{fig2}
\centering
\includegraphics[width=6cm,clip]{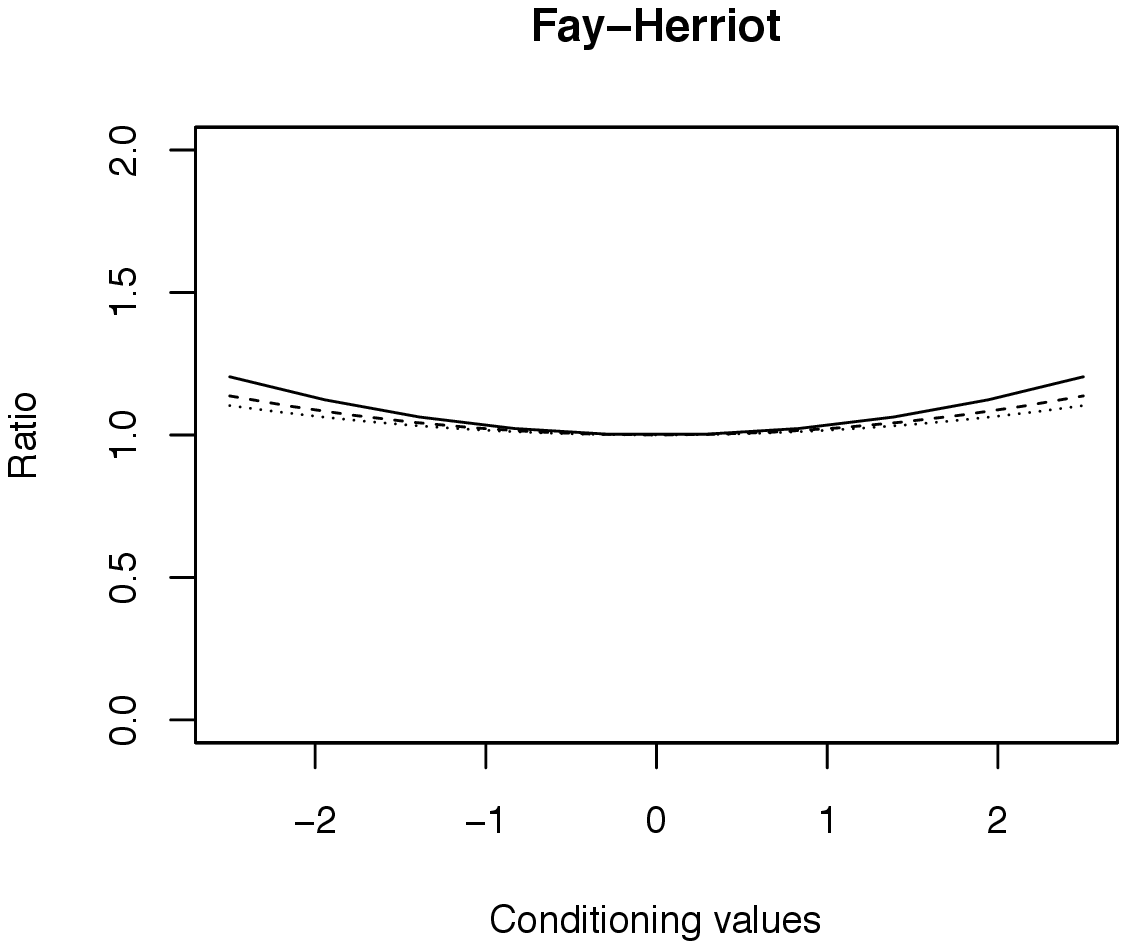}
\includegraphics[width=6cm,clip]{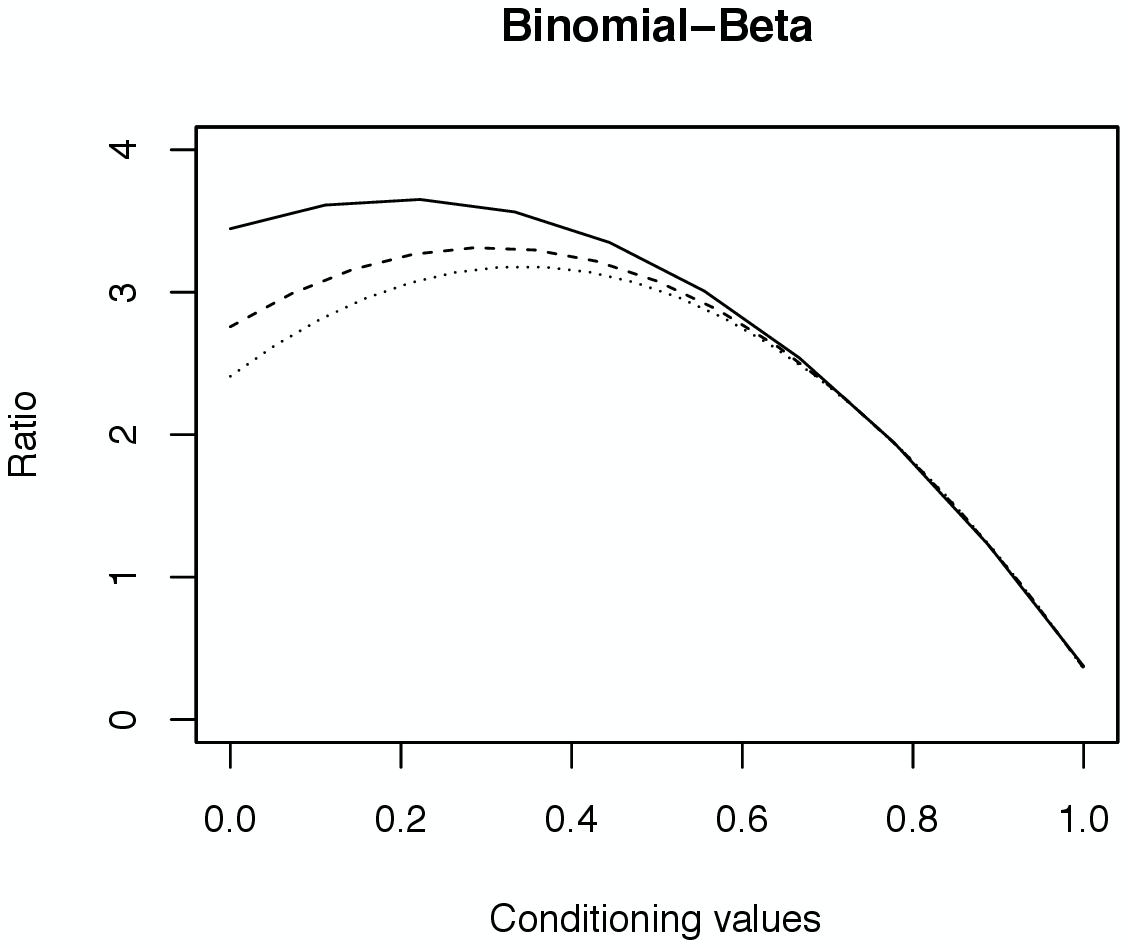}
\includegraphics[width=6cm,clip]{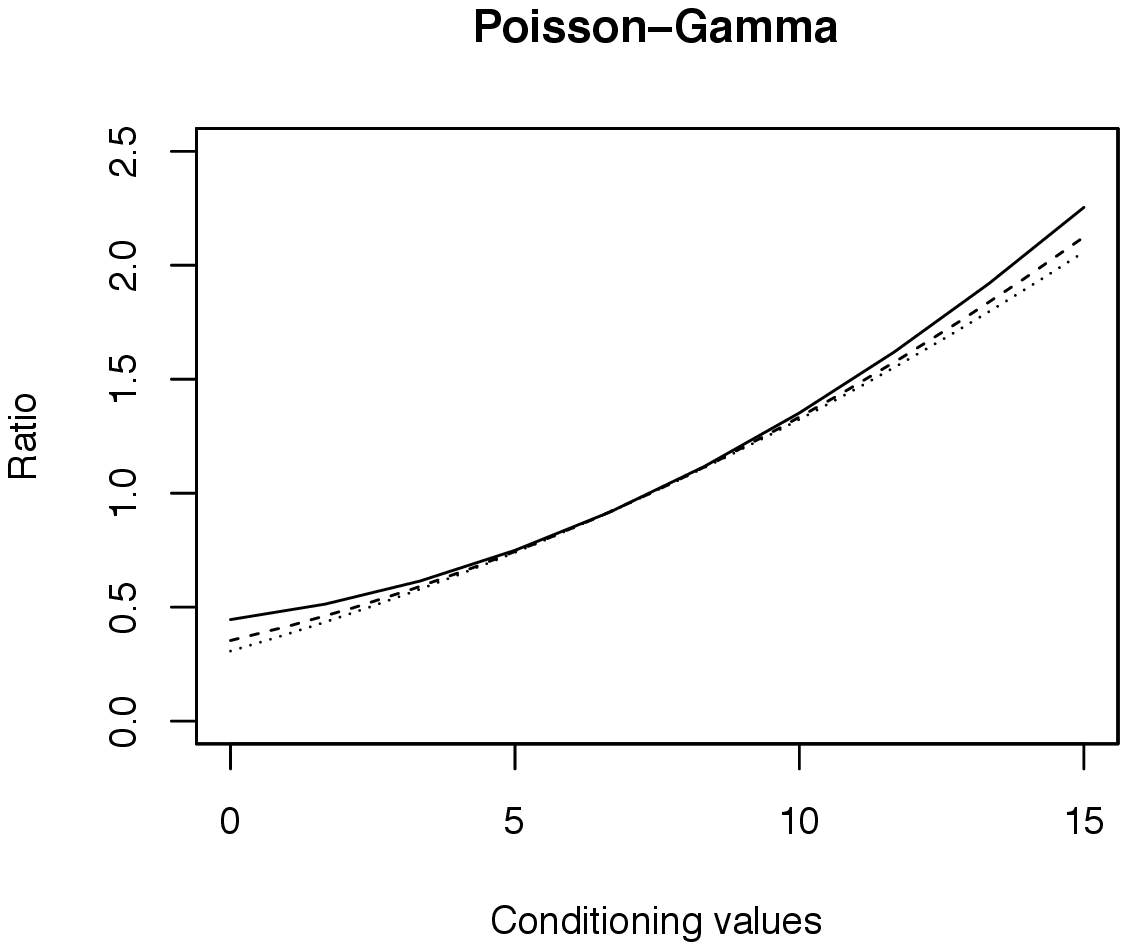}
\caption{Figures of ${\rm Ratio}_2$ for the Fay-Herriot Model (Upper left), the Binomial-beta Mixture Model (Upper right) and the Poisson-gamma Mixture Model (Lower)\   (The solid, dashed and dotted lines correspond to the cases of $m=10$, $15$ and $20$, respectively. The conditioning value denotes $y_i$.)
}
\label{fig:comp2}
\end{figure}

We next investigate the corresponding ratios based on the second-order approximations of the CMSE and MSE. 
Let us define ${\rm Ratio}_2$ by
$$
{\rm Ratio}_2=\{T_{1i}(y_i,\bta)+T_{2i}(y_i,\bta)\}/E[T_{1i}(y_i,\bta)+T_{2i}(y_i,\bta)],
$$
where $T_{1i}(y_i,\bta)+T_{2i}(y_i,\bta)$ and $E[T_{1i}(y_i,\bta)+T_{2i}(y_i,\bta)]$ are given in (\ref{CMSE}) and (\ref{uMSE}), respectively.
Since the second-order terms depend on $m$, we treat the three cases of $m=10$, $15$ and $20$ for $\x_i'\bbe=\mu$ and $n_1=\cdots=n_m=5$. We used $\hbta_{GT}$ for estimation of $\bta$.
The performances of ${\rm Ratio}_2$ are illustrated in Figure \ref{fig:comp2} for the three mixed models, where the values of $(\mu, \nu)$ are $(0, 1)$ for the Fay-Herriot model, $(\exp(2), 1)$ for the Poisson-gamma mixture model, and $(\exp(1.5)/(1+\exp(1.5)), 1)$ for the binomial-beta mixture models.
Figure \ref{fig:comp2} demonstrates that the second-order terms for the three mixed models do not contribute so much to ${\rm Ratio}_2$ or the conditional MSE.

\subsection{Finite performances of the estimator of CMSE}
We investigate finite performances of the second-order unbiased estimator for the conditional MSE by simulation.
The mixed models we examine are the Poisson-gamma mixture and binomial-beta mixture models where the simple case of $\x_i'\bbe=0$ without covariates is treated with $m=25$, $n_i=10$ and $\nu=15$.

In the experiment of simulation, let us fix the index of the area of interest as $i=1$, namely the first area is of interest, and the value of $y_1$ is conditioned.
As seen from the discussion given in Section \ref{sec:com}, the performances of the  conditional MSE depend on the value of $y_1$.
In this simulation, we consider the $\al$-quantile point, denoted by $y_{1(\al)}$, of the distribution of $y_1$ and select the five quantiles $y_{1(\al)}$ for $\alpha=0.05$, $0.25$, $0.5$, $0.75$ and $0.95$.
For the Poisson-gamma mixture model, the marginal distribution of $y_1$ is the negative binomial distribution $NB(\nu m_1,\nu/(n_1+\nu))$, and we can obtain the five quantiles  $y_{1(\al)}$ from the marginal distribution.
For the binomial-beta mixture model, the marginal distribution of $y_1$ is not given as a typical distribution.
Thus, we need to calculate numerically $\alpha$-quantile values of $y_1$.

The true values of CMSE can be provided based on the simulation with $R=10,000$ replications.
For $r=1, \ldots, R$, we generate random variables $y_i^{(r)}$ and $\theta_i^{(r)}$, $i=2,\ldots, m$, which are distributed as $y_i^{(r)}|(\theta_i^{(r)},\mu,\nu)\sim f(y_i|\theta_i^{(r)},\mu,\nu)$ and $\theta_i^{(r)}|(\mu,\nu)\sim \pi(\theta_i|\mu,\nu)$.
In the $r$-th replication, from the sample $\{y_{1(\al)},y_2^{(r)},\ldots,y_m^{(r)}\}$, we calculate the values of $\hxi_1(y_{1(\alpha)},\hbta)^{(r)}$ and $\hxi_1(y_{1(\alpha)},\bta)^{(r)}$. 
Then, the true value of the CMSE of $\hxi_1(y_{1(\alpha)},\hbta)$ can be numerically calculated as
$$
{\rm CMSE}_1=T_{11}(y_{1(\alpha)},\bta)+\frac1R\sum_{r=1}^R\Bigl\{\hxi_1(y_{1(\alpha)},\hbta)^{(r)}-\hxi_1(y_{1(\alpha)},\bta)^{(r)}\Bigr\}^2.
$$
For estimation of the hyperparameter $\bta$, we consider two types of estimators, GT-estimators obtained from estimating equation (\ref{ee}) and ML-estimators by maximizing the marginal likelihood. We used the parametric bootstrap method given in Theorem \ref{thm:cMSEu2}.

Through the same manner as described above, we generate another simulated sample with size $T=2,000$ and calculate the CMSE estimate $\widehat{\rm CMSE}_1$ from (\ref{estcmse}).
Then, we can obtain the relative bias (RB) and coefficients of variation (CV) for the CMSE estimator, which are defined by
\begin{align*}
{\rm RB}=&\frac{T^{-1}\sum_{t=1}^T\widehat{\rm CMSE}_1^{(t)}-{\rm CMSE}_1}{{\rm CMSE}_1},
\\
{\rm CV}=&\Bigl[\frac{1}{T}\sum_{t=1}^T\left(\widehat{\rm CMSE}_1^{(t)}-{\rm CMSE}_1\right)^2\Bigr]^{1/2}\bigg/{\rm CMSE}_1,
\end{align*}
where $\widehat{\rm CMSE}_1^{(t)}$ denotes the CMSE estimate in the $t$-th replication for $t=1, \ldots, T$.

For $\al=0.05$, $0.25$, $0.50$, $0.75$ and $0.95$, the values of $y_{1(\al)}$, ${\rm CMSE}_1$, RB and CV for both GT and ML are reported in Table \ref{tab:sim} for the two mixed models, where the values of ${\rm CMSE}_1$ are multiplied by $100$. 
Table \ref{tab:sim} demonstrates that the estimator $\widehat{{\rm CMSE}}_1^{\rm GT}$ of the conditional MSE performs well for various values of $y_{1(\al)}$ in both models. For $\widehat{{\rm CMSE}}_1^{\rm ML}$, it is biased than $\widehat{{\rm CMSE}}_1^{\rm GT}$, but the CV$^{\rm ML}$ is smaller than CV$^{\rm GT}$, namely the $\widehat{{\rm CMSE}}_1^{\rm ML}$ gives stable estimates. 
The true value of ${\rm CMSE}_i$ has a general trend of increase in $y_{1(\al)}$ for the Poisson-gamma mixture model, and this coincides with the analytical property discussed in Section \ref{sec:com}.
For the binomial-beta mixture model, the true values of ${\rm CMSE}_i$ are about the same and do not have a feature of concavity explained in Section \ref{sec:com}.
The values of RB and CV show that the analytical CMSE estimator based on the GT-estimator and bootstrap CMSE estimator based on the ML-estimator are not bad as an estimator of ${\rm CMSE}_i$.

\small
\begin{table}
\caption{Values of {\rm CMSE}$_1$, Relative Bias {\rm ({RB})} and Coefficient of Variation {\rm  ({CV})} of the CMSE Estimator for the Five Conditioning Values in the Poisson-gamma and Binomial-beta Mixture Models}
\small
\begin{center}
$
{\renewcommand\arraystretch{1.1}\small
\begin{array}{c@{\hspace{3mm}}c@{\hspace{4mm}}
c@{\hspace{3mm}}c@{\hspace{3mm}}c@{\hspace{3mm}}
c@{\hspace{2mm}}c@{\hspace{2mm}}c@{\hspace{2mm}}
c@{\hspace{2mm}}c@{\hspace{2mm}}c@{\hspace{2mm}}
c@{\hspace{2mm}}c@{\hspace{2mm}}c@{\hspace{2mm}}
}
\hline
  & \text{$\alpha$}     & \text{$y_{1(\alpha)}$}   &  \text{CMSE$_1^{\rm GT}$}  & \text{RB$^{\rm GT}$}    & \text{CV$^{\rm GT}$} &  \text{CMSE$_1^{\rm ML}$}  & \text{RB$^{\rm ML}$}    &  \text{CV$^{\rm ML}$}  \\
\hline
&0.05 &0.40  &4.10    &0.09  &0.73  &3.92  &-0.14  &0.20\\
&0.25 &0.70  &3.80    &0.02  &0.53  &3.97  &-0.30  &0.36\\
\text{Poisson-gamma} 
&0.50   &1.00 &4.24   &-0.03  &0.68 &4.31 &-0.36   &0.41\\
&0.75   &1.30 &4.90    &0.05 &0.71 &5.05 &-0.30  &0.37\\
&0.95   &1.70 &6.16    &0.06  &0.66 &6.45 &-0.04  &0.22\\
\hline
&0.05    &0.10   &1.18    &-0.10  &0.30  &1.25  &-0.05 &0.15\\
&0.25    &0.30   &1.07    &0.03  &0.47   &1.10  &-0.24  &0.30\\
\text{Binomial-beta} 
&0.50    &0.40   &1.03    &0.07  &0.56   &1.05  &-0.32 &0.37\\
&0.75    &0.50   &1.03   &0.06 &0.60     &1.03  &-0.34 &0.39\\
&0.95    &0.70   &1.06   &-0.02 &0.51    &1.10  &-0.23 &0.30\\
\hline
\end{array}
}
$
\end{center}
\label{tab:sim}
\end{table}
\normalsize

\subsection{Empirical examples}
\label{sec:EE}

We now apply the suggested procedures to the two data sets: the Stomach Cancer Mortality Data and the Infant Mortality Data Before World War II, both of which are data from prefectures in Japan. In this subsection, we use the analytical CMSE estimator based on the GT-estimator. 

\begin{exm}

{\rm
{\bf  (Mortality rates estimates in the Poisson-gamma mixture model).}\ \ 
We begin by analyzing the Stomach Cancer Mortality Data in Japan.
The data set consists of the observed number of mortality $z_i$ and its expected number $n_i$ of stomach cancer for women who lived in the $i$-th city or town in Saitama prefecture, Japan, for five years from 1995 to 1999. 
Such area-level data $(z_i, n_i)$, $i=1, \ldots, m$, are available for $m=92$ cities and towns, and the total number of mortality in the whole region is $L=3953$. 
The expected numbers are adjusted by age on the basis of the population so that $L=\sum_{i=1}^mz_i=\sum_{i=1}^mn_i$.

For $z_1,\ldots,z_m$, we use the Poisson-gamma mixture model discussed in Section \ref{sec:exm}, namely $z_i|\la_i\sim {\rm Po}(n_i\la_i)$ and $\la_i\sim{\rm Ga}(\nu m_i,1/\nu)$. 
Since data of mortality rate of stomach cancer for men are also available, we can use them as a covariate. 
Let $x_i$ be a log-transformed mortality rate for men for $i$-th area. 
Then, we treat the regression model $\log m_i=\beta_0+x_i\beta_1$ for $i=1,\ldots,m$. 
The unknown parameters $\bta^t=(\beta_0,\beta_1,\nu)^t$ are estimated as the roots of the estimating equations in (\ref{ee}). 
Their estimates are $\beta_0=-7.77\times 10^{-3}, \ \beta_1=0.157$ and $\nu=158$.

To illustrate the difference between CMSE and MSE, we use the percentage relative difference (RD) defined by
$$
{\rm RD}_i=100\times (\widehat{\rm CMSE}_i - \widehat{\rm MSE}_i)/ \widehat{\rm MSE}_i.
$$ 
When ${\rm RD}_i$ is positive, $\widehat{\rm CMSE}_i$ is larger than $\widehat{\rm MSE}_i$.
In Figure \ref{fig:emp1}, the plots of the values $(\widehat{\rm MSE}_i, \widehat{\rm CMSE}_i)$ multiplied by $1,000$ and the values of $(y_i, {\rm RD}_i)$ for $i=1, \ldots, m$ are given in the left and right figures, respectively, where $y_i=z_i/n_i$ is the standard mortality rate (SMR).
From Figure \ref{fig:emp1}, it is revealed that the values of $\widehat{\rm CMSE}_i$ are larger than those of $\widehat{\rm MSE}_i$ for some areas, and that the relative differences ${\rm RD}_i$ have great variability, which comes from non-normality of distribution as discussed in Section \ref{sec:com}.

Table \ref{tab:emp1} reports the values of $n_i$, $y_i$, ${\rm EB}_i$, $\widehat{{\rm CMSE}}_i$, $\widehat{{\rm MSE}}_i$ and ${\rm RD}_i$ for ten selected municipalities in Saitama prefecture, where the values of $\widehat{{\rm MSE}}_i$ and $\widehat{{\rm CMSE}}_i$ are multiplied by $1,000$.  
It is noted that Kumagaya has the maximum RD value and Yoshida has the minimum RD value in our result. 
The values of RD tell us about important information that the given empirical Bayes estimate has a different prediction error from the usual unconditional MSE.
For instance, in Yoshida, the estimate of the CMSE is $8.631$, while that of the unconditional MSE is $18.858$, and the resulting RD is $-54$.
This means that the unconditional MSE over-estimates the CMSE.
On the other hand, in Kumagaya, the estimate of the CMSE is $7.384$, while that of the unconditional MSE is $5.819$, and the resulting RD is $27$.
This means that the unconditional MSE under-estimates the CMSE.
Remember that the CMSE is a function of both $y_i$ and $n_i$ increasing for $y_i$ and decreasing for $n_i$ in the Poisson-gamma model, while the unconditional MSE does not depend on $y_i$ and decreases for $n_i$.
Thus, the CMSE is not always small in areas with small $n_i$ such as Yoshida and Naguri, and the unconditional MSE may over-estimates the CMSE.
On the contrary, in area with large $n_i$ such as Kumagaya, the unconditional MSE may under-estimates the CMSE, which leads to a serious situation in real application.

\begin{figure}
\centering
\includegraphics[width=6cm,clip]{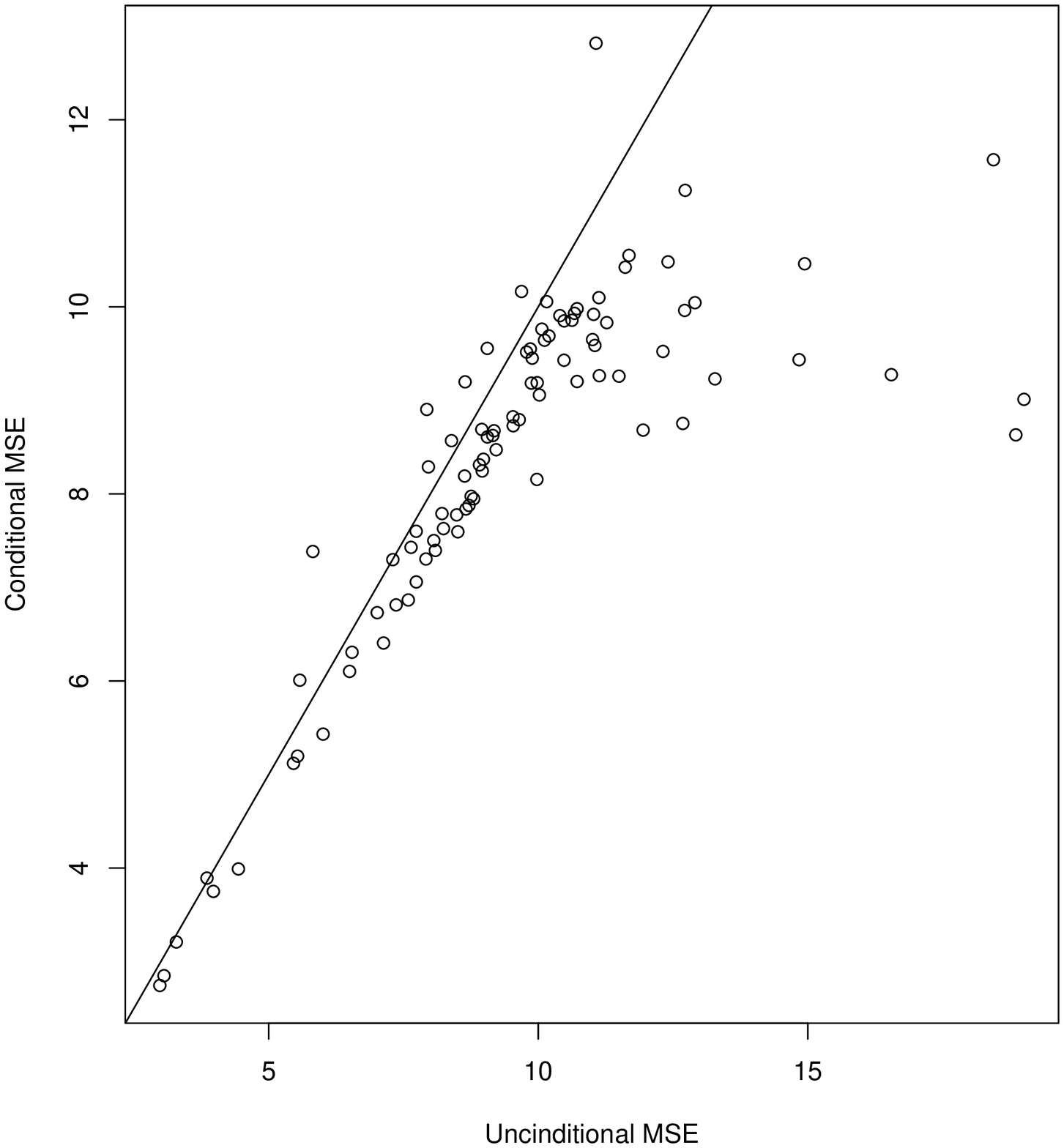}
\includegraphics[width=6cm,clip]{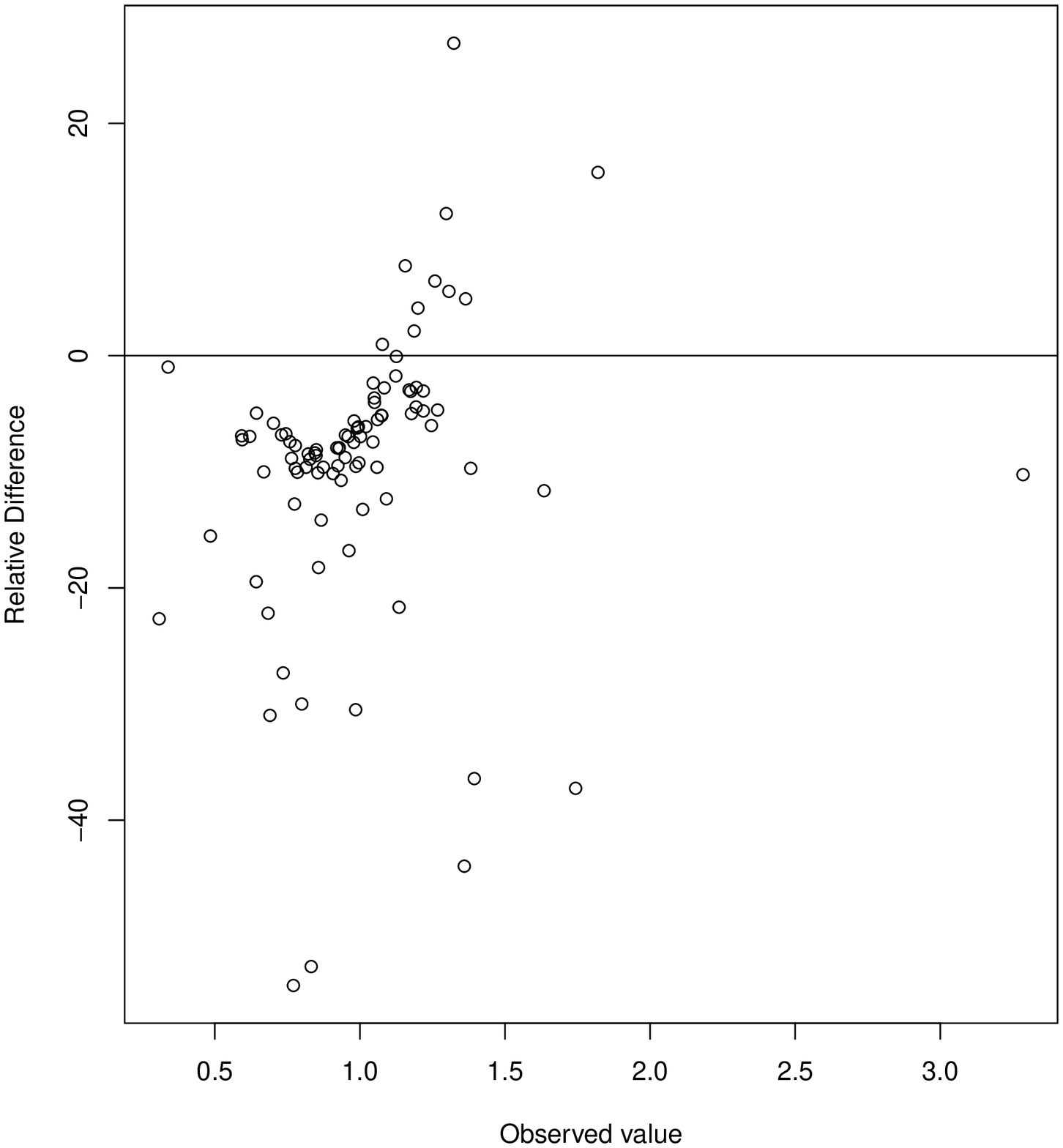}
\caption{Plots  of $(\widehat{\rm MSE}_i, \widehat{\rm CMSE}_i)$ (left) and Plots of $(y_i, {\rm RD}_i)$ (right) for Stomach Cancer Mortality Data}
\label{fig:emp1}
\end{figure}

\small
\begin{table}
\caption{Values of $n_i$, SMR $y_i$, ${\rm EB}_i$, $\widehat{{\rm CMSE}}_i$, $\widehat{{\rm MSE}}_i$ and ${\rm RD}_i$ for Selected Areas in Saitama Prefecture}
\small
\begin{center}
$
{\renewcommand\arraystretch{1.1}\small
\begin{array}{c@{\hspace{3mm}}c@{\hspace{4mm}}
c@{\hspace{3mm}}c@{\hspace{3mm}}c@{\hspace{3mm}}
c@{\hspace{2mm}}c@{\hspace{2mm}}c@{\hspace{2mm}}
c@{\hspace{2mm}}c@{\hspace{2mm}}c@{\hspace{2mm}}
c@{\hspace{2mm}}c@{\hspace{2mm}}c@{\hspace{2mm}}
}
\hline
\text{Area} & \text{$n_i$}     & \ \ \text{$y_i$} \ \ & \text{${\rm EB}_i$}  &  \text{$\widehat{{\rm CMSE}}_i$}  &  \text{$\widehat{{\rm MSE}}_i$}    &  \text{${\rm RD}_i$}\\
\hline
\text{Kawagoe}&192.1 & 1.077 & 1.058 & 3.892 & 3.855 &  1 \\
\text{Kumagaya}&102.7 &1.324  &1.194 & 7.384 & 5.819 & 27\\
\text{Hatagaya}& 35.2 & 1.307 &1.114 & 9.556  & 9.054    &6\\
 \text{Asaka} &52.5 &1.124  &1.031  & 7.600 &  7.736  & -2\\
\text{Sakado}  & 51.6 & 1.298 & 1.131 &  8.903  & 7.933 &  12\\
 \text{Ooi}  & 20.7 & 0.867  &1.003  & 9.202  &10.720 & -14\\
\text{Naguri}   &  3.6 & 1.394 & 0.934 &  9.435 & 14.839 & -36\\
  \text{Yoshida}  & 6.5 & 0.771 &0.863 & 8.631& 18.858 &-54\\
 \text{Kamisato} & 18.3  &1.364  &1.066 & 10.164 &  9.690 &  5\\
 \text{Miyashiro}& 20.1 & 1.194  &1.051  & 9.516   &9.784   &-3\\
\hline
\end{array}
}
$
\end{center}
\label{tab:emp1}
\end{table}
\normalsize

}
\end{exm}

\begin{exm}

{\rm
{\bf (Infant mortality rates estimates in the binomial-beta mixture model).}\ \ 
We next handle the historical data of the Infant Mortality Data Before World War II. 
The data set consists of the observed number of infant mortality $z_i$ and the number of birth $n_i$ in the $i$-th city or town in Ishikawa prefecture, Japan, before World War II. 
Such area-level data are available for $m=211$ cities, towns and villages, and the total number of infant mortality in the whole region is $L=4252$.

It is noted that the infant mortality rates $y_i=z_i/n_i$ before World War II are not small and distributed around $0.2$.
Thus, we here apply the data to the binomial-beta model rather than the Poisson-gamma model.
For $z_1,\ldots,z_m$,  $z_i| p_i$ and $p_i$ have the distributions $z_i|p_i\sim {\rm Bin}(n_i,p_i)$ and $p_i\sim{\rm Beta}(\nu m_i,\nu(1-m_i))$, where $m_i=\exp(\beta)/(1+\exp(\beta))$ for $i=1,\ldots,m$, since we do not have any covariates.
Thus, the unknown parameters are $\bta=(\beta,\nu)^t$ and their estimates are $\beta=-1.57$, namely $m_i=0.171$, and $\nu=102$.

The plots of the values $(\widehat{\rm MSE}_i, \widehat{\rm CMSE}_i)$ multiplied by $1,000$ and the values of $(y_i, {\rm RD}_i)$ for $i=1, \ldots, m$ are given in the left and right figures of Figure \ref{fig:emp2}, respectively.
Figure \ref{fig:emp2} suggests that the values of the relative difference RD increases in $y_i$.
This is because the leading $O_p(1)$ term is an increasing function of $y_i$ for fixed $n_i$ since $y_i$ is between $0$ and $0.5$, as investigated in Section \ref{sec:com}. 
It is observed from Figure \ref{fig:emp2} that the unconditional MSE under-estimates the CMSE in most areas.
This gives us a warning message on the empirical Bayes estimates in each area since the unconditional MSE underestimates the estimation error of the empirical Bayes estimate based on given area data.
Table \ref{tab:emp2} reports the values of $n_i$, $y_i$, ${\rm EB}_i$, $\widehat{{\rm CMSE}}_i$, $\widehat{{\rm MSE}}_i$ and ${\rm RD}_i$ for fifteen selected municipalities in Ishikawa prefecture, where the values of $\widehat{{\rm MSE}}_i$ and $\widehat{{\rm CMSE}}_i$ are multiplied by $1,000$.  
It is noted that Area 175 has the maximum RD value and Area 46 has the minimum RD value in our result. 
For Area 176, the observed mortality rate $y_i=0.400$ is much shrunken to ${\rm EB}_i=0.216$ by the empirical Bayes estimator since the number of birth is quite small as given by $n_i=25$.
The unconditional MSE is estimated by $1.216$, but the relative difference is ${\rm RD}_i=62$, and the estimate of CMSE is $1.964$, which is higher than the MSE estimate.
This suggests that it should be good to provide estimates of CMSE as well as estimates of MSE.

\begin{figure}[t]
\centering
\includegraphics[width=6cm,clip]{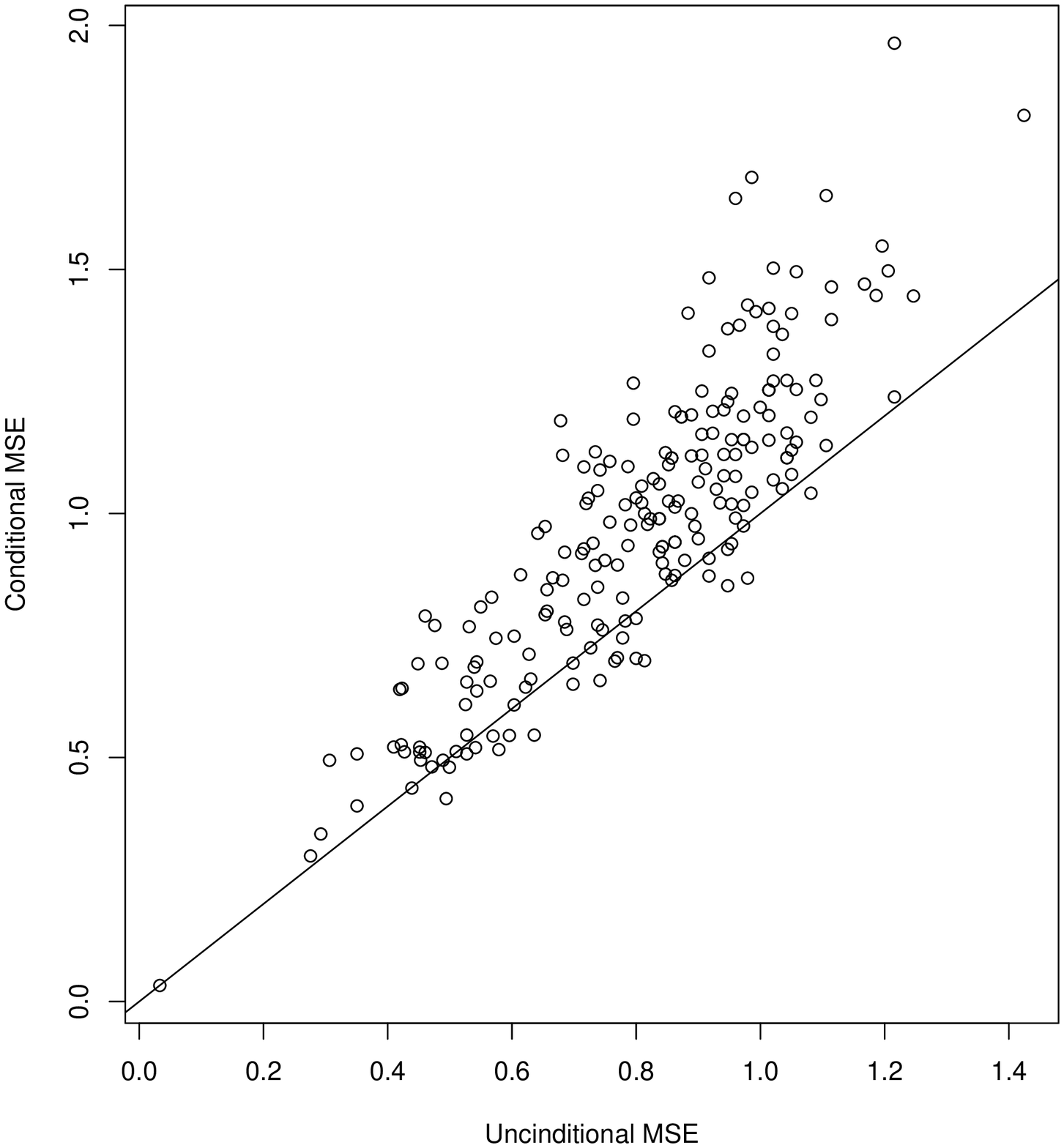}
\includegraphics[width=6cm,clip]{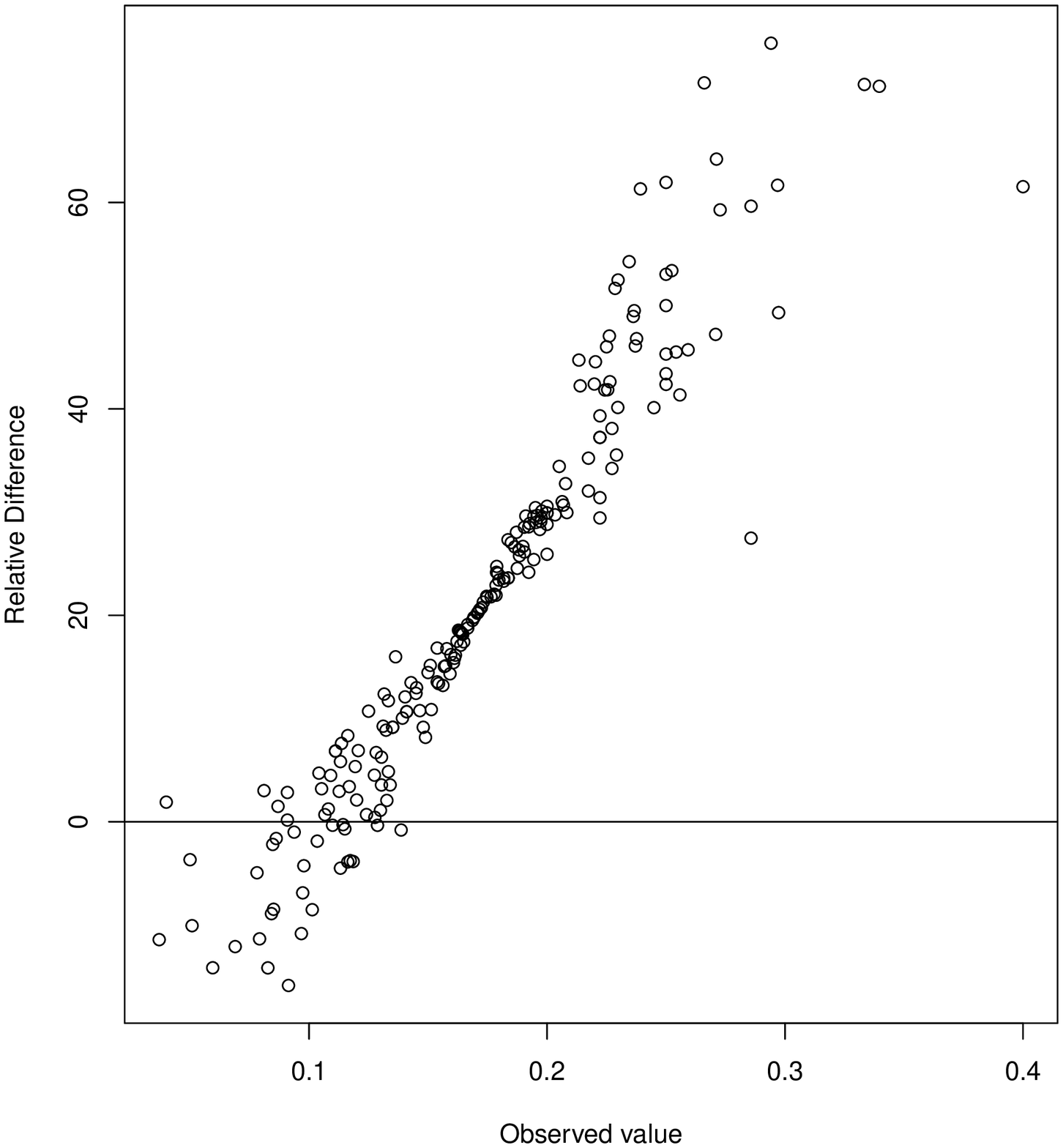}
\caption{Plots  of $(\widehat{\rm MSE}_i, \widehat{\rm CMSE}_i)$ (left) and Plots of $(y_i, {\rm RD}_i)$ (right) for Infant Mortality Data}
\label{fig:emp2}
\end{figure}

\small
\begin{table}
\caption{Values of $n_i$, $y_i$, ${\rm EB}_i$, $\widehat{{\rm CMSE}}_i$, $\widehat{{\rm MSE}}_i$ and ${\rm RD}_i$ for Selected Areas in Ishikawa Prefecture}
\small
\begin{center}
$
{\renewcommand\arraystretch{1.1}\small
\begin{array}{c@{\hspace{3mm}}c@{\hspace{4mm}}
c@{\hspace{3mm}}c@{\hspace{3mm}}c@{\hspace{3mm}}
c@{\hspace{2mm}}c@{\hspace{2mm}}c@{\hspace{2mm}}
c@{\hspace{2mm}}c@{\hspace{2mm}}c@{\hspace{2mm}}
c@{\hspace{2mm}}c@{\hspace{2mm}}c@{\hspace{2mm}}
}
\hline
\text{Area} & \text{$n_i$}     & \ \ \text{$y_i$} \ \ & \text{${\rm EB}_i$}  &  \text{$\widehat{{\rm CMSE}}_i$}  &  \text{$\widehat{{\rm MSE}}_i$}    &  \text{${\rm RD}_i$}\\
\hline
1         &  4146  & 0.139 &  0.139 & 0.033 & 0.033 & 0\\
19     & 56        &0.250  & 0.199  &1.386  &0.966  & 43  \\
23     &  55       & 0.164 & 0.168  & 1.152  & 0.973  &  18\\
46      & 197     &0.091  & 0.119  & 0.416  & 0.494  &  -16\\
71      &84        &0.060   & 0.121  & 0.698  & 0.814  & -14\\
79      & 87       &0.069   & 0.124  & 0.703 & 0.800  & -13\\
86      &101       &0.079  & 0.125  & 0.658 & 0.742 & -11\\
96      &194       &0.119  & 0.137  &0.480  & 0.499  & -4\\
98     &  208     &0.250  & 0.224  & 0.771 & 0.476  & 62\\
112  &  94        & 0.160 & 0.166  & 0.894  & 0.770 & 16\\
158    &173      & 0.185  & 0.180  & 0.685  & 0.539 &27\\
162    & 57        &0.333   & 0.229 & 1.646 & 0.960 & 71\\
175    &119      & 0.294  & 0.237  &1.190  & 0.678 & 75\\
176    & 25        & 0.400  & 0.216 & 1.964 & 1.216 &  62\\
179  &  245     & 0.229  & 0.212 & 0.642  & 0.423 &  52\\
\hline
\end{array}
}
$
\end{center}
\label{tab:emp2}
\end{table}
\normalsize

}
\end{exm}

\section{Concluding Remarks}
\label{sec:CR}

In this paper, we have derived the second-order approximation of the conditional MSE of the empirical Bayes estimator and its second-order unbiased estimator in the general mixed models.
Those results have been applied to the mixed models based on NEF-QVF, and the second-order evaluations of the CMSE have been provided in analytical and closed forms for the GT-estimator and the parametric bootstrap method for the ML-estimator without assuming that the sample size $n_i$ goes to infinity.
It has been shown that the difference between the conditional and unconditional MSEs is small for the normal distribution, while it is significant for the Poisson-gamma and the binomial-beta mixture models.
We have also clarified how different the CMSE is from the MSE by comparing the leading terms in the CMSE and MSE.

Concerning the two measures for evaluating the estimation error, one important issue is which one should go for the conditional or the unconditional approach.
In general, this issue depends on what one wants to know as the estimation error.
When data of the small area of interest are observed and one wants to know the estimation error of the empirical Bayes estimate based on these data, the CMSE given the data should be used.
When one wants to know the average estimation error of the estimator, on the other hand, the unconditional MSE is employed.
As illustrated in Figure \ref{fig:comp2}, however, the ratio of the CMSE over the unconditional MSE is significant for the Poisson-gamma and the binomial-beta mixture models, while it is close to one for the normal Fay-Herriot model.
The similar discrepancy between the two methods is shown in Figures \ref{fig:emp1} and \ref{fig:emp2}, and Tables \ref{tab:emp1} and \ref{tab:emp2} for the two examples.
For example, in Table \ref{tab:emp2}, the estimated CMSE of area 175 is 1.190, while the estimated unconditional MSE is 0.678.
This gives us a warning message on the value 0.237 of the empirical Bayes estimate based on the data from area 175.
These observations reveal the risk that the unconditional MSE sometimes underestimates the estimation error of the empirical Bayes estimate based on given area data.
Thus, we suggest providing estimates of the CMSE.

\section*{Acknowledgments}
We would like to thank the associate editor and the reviewer for some important comments which led to an improved version of this paper.
The first author was supported in part by Grant-in-Aid for Scientific Research (15J10076) from Japan Society for the Promotion of Science (JSPS).
The second author acknowledges support from Grant-in-Aid for Scientific Research (15H01943 and 26330036), Japan.

\appendix

\medskip
\section{Proof of Lemma \ref{lem:1}}

For notational simplicity, we put $\R_i=\D_i^t\bSi_i^{-1}$ and we use $\U$ as $\U(\bta)$.
Using the results in Ghosh and Maiti (2004), we immediately have $\hbta-\bta=\U^{-1}\s_m+o_p(m^{-1/2})$, which implies that
$$
E\left[(\hbta-\bta)(\hbta-\bta)^t|y_i\right]=\U^{-1}E\left[\s_m\s_m^t|y_i\right]\U^{-1}+o_p(m^{-1}),
$$
where
\begin{align*}
E\left[\s_m\s_m^t|y_i\right]&=\sum_{j=1}^mE\left[\R_j\g_j\g_j^t\R_j^t|y_i\right]=\sum_{j\neq i}^mE\left[\R_j\g_j\g_j^t\R_j^t\right]+\R_i\g_i\g_i^t\R_i^t\\
&=\U+\R_i(\g_i\g_i^t-\bSi_i)\R_i^t,
\end{align*}
since $\g_j$ depends only on $y_j$ of $\Y$ and $y_1,\ldots,y_m$ are mutually independent.
Since $\U=O(m)$ and $\R_i(\g_i\g_i^t-\bSi_i)\R_i^t=O_p(1)$, we have
$E\left[\s_m\s_m^t|y_i\right]=\U+O_p(1)$, so that 
\begin{equation}\label{cond}
E\left[(\hbta-\bta)(\hbta-\bta)^t|y_i\right]=\U^{-1}+o_p(m^{-1}).
\end{equation}

\ \\
Next, we evaluate asymptotically the conditional bias of $\hbta$, i.e. $E[\hbta-\bta|y_i]$. 
Expanding the equation (\ref{ee}) up to second order, we have
\begin{align*}
\hbta-\bta=\Bigl(-\frac{\pd \s_m}{\pd\bta}\Bigr)^{-1}\Bigl(\s_m+\frac12 \t+o_p(1)\Bigr),
\end{align*}
where
$$
\frac{\pd\s_m}{\pd\bta^t}=\sum_{j=1}^m\Bigl(\frac{\pd\R_j }{\pd\bta^t}\Bigr)\left(\I_p\otimes\g_j\right)+\sum_{j=1}^m\R_j\Bigl(\frac{\pd\g_j}{\pd\bta^t}\Bigr),
$$
noting that $\pd\s_m/\pd\bta^t=-\U+o_p(m)$, and
$$
\t=\col_{\ell}\Bigl\{(\hbta-\bta)^t\left(\frac{\pd^2 S_{m\ell}}{\pd\bta\pd\bta^t}\right)(\hbta-\bta)\Bigr\},
$$
for $\s_m=(S_{m1},\ldots,S_{mq})$ with $q=p+1$.
It noted that $S_{mk}=\R_{ik}\g_i$ for $k=1,\ldots,q$, where $\R_{ik}$ is the $k$-th row vector of $\R_i$.
The notation $\col_{\ell}\left\{a_{\ell}\right\}$ for scalars $a_{\ell}$'s, $\ell=1,\ldots,n$ is defined by
$$
\col_{\ell}\left\{a_{\ell}\right\}=(a_1, a_2, \ldots, a_n)^t.
$$
Let $\W=\pd\s_m/\pd\bta^t-(-\U)$, then we have
$$
\Bigl(-\frac{\pd \s_m}{\pd\bta}\Bigr)^{-1}=-\U^{-1}-\U^{-1}\W\U^{-1}+o_p(m^{-3/2}).
$$
Therefore, it follows that
\begin{align*}
\btah-\bta&=\left(\U^{-1}+\U^{-1}\W\U^{-1}+o_p(m^{-3/2})\right)\left(\s_m+\frac12\t+o_p(1)\right)\\
&=\U^{-1}\s_m+\frac12\U^{-1}\t+\U^{-1}\W\U^{-1}\s_m+o_p(m^{-1}),
\end{align*}
whereby
\begin{align}\label{cb}
E[\btah-\bta|y_i]=\U^{-1}\R_i\g_i+\frac12\U^{-1}E\left[\t|y_i\right]+\U^{-1}E\left[\W\U^{-1}\s_m|y_i\right].
\end{align}
For the second term in (\ref{cb}), note that
\begin{align*}
E\left[\t|y_i\right]&=\col_{\ell}\Bigl\{E\left[(\hbta-\bta)^t\left(\frac{\pd^2 S_{m\ell}}{\pd\bta\pd\bta^t}\right)(\hbta-\bta)\bigg|y_i\right]\Bigr\}\\
&=\col_{\ell}\Bigl\{\tr\Bigl\{ \left(\frac{\pd^2 S_{m\ell}}{\pd\bta\pd\bta^t}\right) E\left[(\hbta-\bta)^t(\hbta-\bta)\bigg|y_i\right]\Bigr\}\Bigr\}\\
&=\col_{\ell}\Bigl\{\tr\Bigl(E\Bigl[\frac{\pd^2 S_{m\ell}}{\pd\bta\pd\bta^t}\Bigr]\U^{-1}\Bigr)\Bigr\}+o_p(1)\equiv \a_2(\bta)+o_p(1).
\end{align*}
The straightforward calculation shows that
$$
\frac{\pd^2 S_{m\ell}}{\pd\bta\pd\bta^t}=\sum_{i=1}^m\left\{\left(\frac{\pd^2 \R_{i\ell}}{\pd\bta\pd\bta^t}\right)(\I_p\otimes \g_i)+2\frac{\pd \R_{i\ell}}{\pd\bta}\frac{\pd \g_i}{\pd\bta^t}+(\I_p\otimes \R_{i\ell})\left(\frac{\pd^2 \g_i}{\pd\bta^t\pd\bta}\right)\right\},
$$
so that 
$$
E\left[\frac{\pd^2 S_{m\ell}}{\pd\bta\pd\bta^t}\right]=\sum_{i=1}^m\left\{2\left(\frac{\pd \R_{i\ell}}{\pd\bta}\right)\D_i+(\I_p\otimes \R_{i\ell})E\left(\frac{\pd^2 \g_i}{\pd\bta^t\pd\bta}\right)\right\}.
$$
Since 
\begin{equation}\label{dg}
\frac{\pd \g_i}{\pd\bta^t}=2Q(m_i)(y_i-m_i)
\Bigl(\begin{array}{cc}
\0^t & 0\\
\x_i^t & 0
\end{array}\Bigr)
-\D_i,
\end{equation}
we obtain
\begin{align*}
\frac{\pd^2 \g_i}{\pd\bta^t\pd\bta}=
\Bigl(\begin{array}{c}
2\x_i Q(m_i)\left\{Q'(m_i)(y_i-m_i)-Q(m_i)\right\} \\
0
\end{array}\Bigr)
\otimes
\Bigl(\begin{array}{cc}
\0^t & 0\\
\x_i^t & 0
\end{array}\Bigr)
-\frac{\pd\D_i}{\pd\bta},
\end{align*}
whereby
$$
\Z_i\equiv E\left(\frac{\pd^2 \g_i}{\pd\bta\pd\bta^t}\right)=
\Bigl(\begin{array}{c}
-2\x_i Q(m_i)^2\\
0
\end{array}\Bigr)
\otimes
\Bigl(\begin{array}{cc}
\0 & \x_i\\
0 & 0
\end{array}\Bigr)
-\frac{\pd\D_i}{\pd\bta}.
$$
Then we have
\begin{equation}\label{a2}
\a_2(\bta)=\col_{\ell}\Bigl\{\tr\Bigl(\U^{-1}\sum_{i=1}^m\Bigl\{2\Bigl(\frac{\pd \R_{i\ell}}{\pd\bta}\Bigr)\D_i+(\I_q\otimes \R_{i\ell})\Z_i\Bigr\}\Bigr)\Bigr\}.
\end{equation}
For the evaluation of the third term in (\ref{cb}), we get
$$
\U^{-1}E\left[\W\U^{-1}\s_m|y_i\right]=\U^{-1}E\left[\W\U^{-1}\s_m\right]+o_p(m^{-1}),
$$
and
\begin{align*}
E&\left[\W\U^{-1}\s_m\right]=E\left[\left(\frac{\pd \s_m}{\pd\bta^t}\right)\U^{-1}\s_m\right]\\
&=\sum_{i=1}^m\left(\frac{\pd \R_i}{\pd\bta^t}\right)E\left[(\I_p\otimes \g_i)\U^{-1}\R_i\g_i\right]+\sum_{i=1}^m\R_{i}E\left[\left(\frac{\pd \g_i}{\pd\bta^t}\right)\U^{-1}\R_i\g_i\right].
\end{align*}
Using the expression (\ref{dg}), we finally have
\begin{equation}\label{a1}
\begin{split}
\a_1(\bta)&\equiv E\left[\W\U^{-1}\s_m\right]\\
&=\sum_{i=1}^m\left(\frac{\pd \R_i}{\pd\bta^t}\right)\vec(\D_i\U^{-1})+2\sum_{i=1}^m\R_{i}Q(m_i)
\Bigl(\begin{array}{cc}
\0^t & 0\\
\x_i^t & 0
\end{array}\Bigr)
\U^{-1}\R_i
\Bigl(\begin{array}{c}
\mu_{2i}\\  \mu_{3i}
\end{array}\Bigr),
\end{split}
\end{equation}
which completes the proof.
\hfill$\Box$

\section{Numerical evaluation of partial derivatives.}

The analytical expression of $\pd\R_i/\pd\bta^t$ and $\pd\D_i/\pd\bta$ are complex and not practical.
However, the values of these derivatives at some value $\bta_0$ can be easily calculated.
Let $z_m$ be a positive number depending on $m$, then the value of $\pd\R_i/\pd\eta_k, \ k=1,\ldots,k$ at $\bta=\bta_0$ is evaluated as
$$
\frac{\pd\R_i}{\pd\eta_k}(\bta_0)^{\ast}\equiv (2z_m)^{-1}\left\{\R_i(\bta_0+z_m\e_k)-\R_i(\bta_0-z_m\e_k)\right\},
$$
where $\e_k$ is a vector of $0$'s other than $k$-th element is $1$.
Since the difference between $\pd\R_i/\pd\eta_k$ and $\pd\R_i/\pd\eta_k^{\ast}$ at $\bta=\bta_0$ is $O(z_m)$, the choice $z_m=o(m^{-1})$ does not affect the second-order unbiasedness of the CMSE estimator established in Theorem \ref{thm:unef}.
In numerical studies given in this paper, we choose $z_m=m^{-5/4}$ satisfying $z_m=o(m^{-1})$.
The partial derivative $\pd\D_i/\pd\bta$ can be numerically evaluated in the same way.

\ \\
\ \\


\begin{thebibliography}{00}
\bibitem{}
Booth, J. S. and Hobert, P. (1998). Standard errors of prediction in generalized linear mixed models. 
{\it J. Amer. Statist. Assoc.}, {\bf 93}, 262-272. 


\bibitem{}
Datta, G.S., Kubokawa, T., Molina, I. and Rao, J.N.K. (2011). Estimation of mean squared error of model-based small area estimators, {\it TEST}, {\bf 20}, 367-388.

\bibitem{} 
Datta, G.S., Rao, J.N.K. and Smith, D.D. (2005).
On measuring the variability of small area estimators under a basic area level model.
{\it Biometrika}, {\bf 92}, 183-196.


\bibitem{}
Fay, R. E. and Herriot, R. A. (1979). Estimates of income for small places: an application of James-Stein procedures to census data. 
{\it J. Amer. Statist. Assoc.}, {\bf74}, 269-277.

\bibitem{}
Godambe, V. P. and Thompson, M. E. (1989). An extension of quasi-likelihood estimation (with Discussion). 
{\it J. Statist. Plan. Inf.}, {\bf22}, 137-152.

\bibitem{}
Ghosh, M. and Maiti, T. (2004). Small-area estimation based on natural exponential family quadratic variance function models and survey weights. 
{\it Biometrika}, {\bf 91}, 95-112.

\bibitem{}
Ghosh, M. and Maiti, T. (2008). Empirical Bayes confidence intervals for means of natural exponential family-quadratic variance function distributions with application to small area estimation. 
{\it Scand. J. Statist.}, {\bf 35}, 484-495.

\bibitem{}
Ghosh, M. and Rao, J.N.K. (1994).
Small area estimation: An appraisal.
{\it Statist. Science}, {\bf 9}, 55-93.


\bibitem{}
Hall, P. and Maiti, T. (2006). On parametric bootstrap methods for small area prediction. 
{\it J. R. Stat. Soc. Ser. B Stat. Methodol.}, {\bf 68}, 221-238.

\bibitem{}
Kubokawa, T., Hasukawa, M. and Takahashi, K. (2014). On measuring uncertainty of benchmarked predictors with application to disease risk estimate. 
{\it Scand. J. Statist.}, {\bf 41}, 394-413.

\bibitem{}
Morris, C. (1982). Natural exponential families with quadratic variance functions. 
 {\it Ann. Statist.}, {\bf 10}, 65-80. 

\bibitem{}
Morris, C. (1983). Natural exponential families with quadratic variance functions: Statistical theory. 
 {\it Ann. Statist.}, {\bf 11}, 515-529.

\bibitem{} 
Prasad, N. and Rao, J. N. K. (1990).
The estimation of mean-squared errors of small-area estimators. 
{\it J. Amer. Statist. Assoc.}, {\bf 90}, 758-766.

\bibitem{}
Rao, J.N.K. (2003).
{\it Small Area Estimation}. Wiley.

\bibitem{}
Torabi, M. and Rao, J.N.K. (2013). Estimation of mean squared error of model-based estimators of small area means under a nested error linear regression model. 
{\it J. Multivariate Anal.}, {\bf 117}, 76-87.

\end{thebibliography}
\end{document}